\documentclass[usegraphicx,usenatbib]{mn2e}
\usepackage{enumerate}
\usepackage{graphics}
\usepackage{color}
\usepackage{mathrsfs}
\usepackage{amsmath}
\usepackage{comment}

\newcommand{\msun}{M_\odot}
\newcommand{\rsun}{R_\odot}

\newcommand{\eg}{\emph{e.g.}}

\newcommand{\gv}[1]{\ensuremath{\mbox{\boldmath$ #1 $}}} 
\newcommand{\grad}[1]{\gv{\nabla} #1}

\begin{document}

\title{Mass transfer from giant donors}

\author[Pavlovskii \& Ivanova]{K.~Pavlovskii,$^1$ N.~Ivanova,$^1$ \\
$^1$University of Alberta, Dept.\ of Physics, 11322-89 Ave, Edmonton, AB, T6G
2E7, Canada}
 
\maketitle
\begin{abstract}
The  stability of  mass  transfer in  binaries  with convective  giant
donors remains  an open  question in modern  astrophysics. There  is a
significant discrepancy between what  the existing methods predict for
a response to mass  loss of the giant itself, as well  as for the mass
transfer rate  during the Roche lobe  overflow. Here we show  that the
recombination energy  in the  superadiabatic layer plays  an important
and  hitherto unaccounted-for  role in  the donor's  response to  mass
loss, in particular  on its luminosity and  effective temperature. Our
improved optically thick nozzle method  to calculate the mass transfer
rate via $L_1$ allows us to  evolve binary systems for a substantial
Roche  lobe overflow. We  propose a new,  strengthened criterion
for the mass transfer instability, basing  it on whether the donor experiences 
 overflow through its outer Lagrangian point.
  We find that with the new criterion, if the donor
has a  well-developed outer convective envelope,  the critical initial
mass  ratio  for  which  a  binary would  evolve  stably  through  the
conservative mass transfer varies from $1.5$ to $2.2$, which is about twice as large
as  previously  believed. In underdeveloped giants with shallow convective envelopes this critical ratio may be even larger.
When  the  convective  envelope is  still growing, and in  particular for most cases of  massive donors, the
critical mass ratio  gradually decreases to this value,  from that of 
radiative donors.
\end{abstract}


\section{Introduction} \label{paper_introduction}

Many interacting binaries start their mass exchange when the donor,
which has evolved to the giant branch, overfills its Roche lobe. The
stability at the start of the mass transfer (MT) differentiates
between the binaries that will live long as a mass transferring system
and will appear, \eg, as an X-ray binary, and ones that will be
transformed dramatically by a common envelope event.  A clear
separation of the possible evolutionary channels is important for our
understanding of the formation of compact binaries. E.g., double
white-dwarf binaries were thought in the past to be formed via two
common envelope events \citep{Nelemans00, Nelemans05}, while more
recent detailed simulations of MT with low-mass red giant donors show
that the first episode of MT was stable \citep{Woods12}.

There are cases when MT from a main sequence star, while deemed to be stable
  initially, increases its rate significantly with time, in some cases
  leading to so-called delayed dynamical instability \citep[for a
    thorough discussion, see][]{Ge10}. A delayed rapid growth of MT rate
  can also occur in systems with giant donors\citep{Woods12}.
  Unfortunately, to date there exists no simple criterion for the
  start of a dynamically unstable MT that would be based just on its
  rate. It can be defined as MT happening on the dynamical timescale of the donor,
  but even this definition does not determine the fate of a system \citep[for a review of various problems with
    the initiation of a common envelope event, see][]{Ivanova12}.

Standard definition of a dynamical-timescale MT is that the mass is transferred
on the donor's dynamical timescale (see, \eg, \citealt{Paczynski1972}).
When a star is subjected to a dynamical-timescale mass loss (ML), stellar codes operate outside of the regime in which they were designed to operate,
leading to poorly understood behaviour.  Indeed, a conventional stellar
code, developed to treat evolutionary changes, typically does not
include hydrodynamical terms in structure equations, and hence
is unable to treat dynamical-timescale features of the ML response.

There are then two ways to deal with a dynamical ML in practice, both
of them relying on plausible-sounding but arbitrary assumptions.

One way is to pre-set a threshold for the MT rate close to (but still
within) the limits of validity of a conventional stellar code, and
assume MT to become unstable once that threshold is exceeded. For
example, a dynamical MT has been  defined as 10 times the thermal
timescale MT \citep[][]{2001ApJ...552..664N}.

Another way is to assume that during a very short period of very rapid mass loss (ML),
any mass shell in the stellar model has no time to exchange heat with its neighbouring shells.
Hence, its entropy remains almost the same as at the start of the ML,
in the other words the specific entropy profile of the donor is ``frozen''.

The MT rate and stability depend crucially on the response of the
donor's radius to the ML compared to the response of its Roche lobe
radius to the same ML \cite[for foundations, see][]{Webbink85}.  The
governing feature that shaped studies of MT in binary systems with
giant donors was the wide acceptance of the theory that a convective
donor would intrinsically expand as a reaction to ML.  In the
framework of this theory, the immediate response of a star to ML takes
place on a dynamical timescale (to restore its hydrostatic
equilibrium) -- this assumes that no thermal effects can take place,
and it was hence called an adiabatic response.  It is characterised by
the quantity known as the mass-radius exponent $\zeta_{\rm ad}$

\begin{equation} 
\zeta_{\rm ad} \equiv \left ( \frac{\partial \log R}{\partial
\log M} \right) _{\rm ad} \ .  
\end{equation}

With polytropic models (including also composite and condensed
polytropes), it was found that a convective donor would have
$\zeta_{\rm{ad}} < 0$ \citep{Paczynski65, Hjellming87}.  The presence
of a non-convective core increases the value of $\zeta_{\rm{ad}}$,
which becomes positive once the relative mass of the core is more than
20\% of the star's total mass \citep{Hjellming87}.

It is commonly assumed that MT is dynamically stable only if a donor
remains within its Roche lobe, or when $\zeta_{\rm L} \le \zeta_{\rm
  ad}$, where $\zeta_{\rm L}$ is the mass-radius exponent for the
Roche lobe reaction to the MT\citep{Webbink1985}.  Using the Roche lobe approximation as
in \cite{1983ApJ...268..368E}, for conservative MT, $\zeta_{\rm L}= 0
$ when the mass ratio is $q_{\rm crit}\approx 0.788$, and is positive
for a larger mass ratio. For completely non-conservative isotropic
re-emission this critical mass ratio is $\approx 1.2$
\citep{Soberman97}.

\cite{Hjellming87} showed that taking into account the core improves
stability, however for all stars with relative mass of the core
$\la 0.45$ of the total stellar mass, the first episode of
conservative MT will be dynamicall unstable.  Later, the results based on
polytropic models were re-evaluated in the studies of the detailed
adiabatic stellar models by \cite{Ge10}. At the same time, detailed codes
that traced mass transfer in binaries up to thermal timescale mass transfer,
found that critical mass ratio can be up to 1.1 \citep[\eg,][]{2002MNRAS.336..449H}.

It is crucial to realize that the difference in rates between a
  nearly-thermal timescale MT that can be legitimately calculated with a
  conventional stellar code and a dynamical-timescale MT -- the regime of
  adiabatic codes -- is several orders of magnitude, and currently
  there is no adequate treatment to describe the donor response in between
  these two regimes.

A detailed stellar code can be supplied with hydrodynamical
terms in its structure equations and be forced to work in a hydrodynamical
regime (please note that adiabatic codes do not include
hydrodynamical terms in them). This way, in the recent study of
responses of stellar models of giants to fast ML, it was found that
this response is a function of the MT rate \citep{Woods2011}.  As a
result, it was found that for a large range of ML rates taking place
in binaries, giants can also contract, and the nature of the donor's
response was attributed to the behaviour of the giant's superadiabatic
surface layer and its short thermal timescale. In that study, two
different stellar codes were used for comparison and they showed the
same results.
The two codes were: a) STARS/ev code developed by Eggleton (\cite{Eggleton71, Eggleton72, Eggleton73}; \cite{Eggletonetal73}) with the most recent update described as in \cite{Glebbeek08};
b) the Heyney-type code developed by \cite{Kippenhahn67} with the most recent update described in \cite{Ivanova04}.
These results were later confirmed by \cite{Passy2012},
with yet another stellar code, {\tt MESA}.

As can be seen, three different detailed stellar codes
appeared to produce the same result -- the
donor response is very different from the expectations given by a
simplified adiabatic model.  However, no further analysis of {\it why}
there is a dramatic difference between an adiabatic approach and a
detailed stellar code was provided.  The important question remains:
which result we should trust more? What could be the possible
problems with the adiabatic approach? Could we trust the results of
the detailed stellar codes when they are forced to work at dynamical-timescale
MT rates, and when they possibly break down? 

It is clear that for further progress in understanding of the
response of the giant's radius to fast ML, one needs to understand
better the non-adiabatic processes taking place in a red giant
subjected to dynamical-timescale ML. This is intrinsically related
to the nature -- the underlying physics and structure -- of the
superadiabatic layer, which is not yet well understood. A possible
influence  of the surface layer in giants on MT was realised in the
past. For example, \cite{1970ApJ...162..621O}  had suggested that,
in U Gem giants, the mechanism for dynamical mass exchange, proposed
by  \cite{Paczynski65}, can not work, ``because of changes
occurring on the thermal  time scales of the photosphere and
transition zone, which are shorter than the  dynamical time scale of
the star'').  However, the physics of the superadiabatic layer had
never been studied in detail.   On the other hand, the difference in
the response in different codes also could  be related to numerical
effects, which could be either inconsistency in the adopted set of
equations, or numerical artefacts.

In this paper, in \S2, we systematically analyse the physics of the
superadiabatic  layer: what difference an adiabatic approach makes,
which artefacts might appear if this layer is not numerically modelled
properly, what affects the structure of this layer, how massive this layer is in
different donors, and finally, whether we can numerically obtain the
response of the stellar radius  properly with a detailed stellar code,
and with what limitations. 

All detailed stellar models and numerical experiments in this paper
are calculated with {\tt MESA} (Modules for Experiments in Stellar
Astrophysics, http://mesa.sourceforge.net).  This set of stellar
libraries is described in \cite{Paxton2011, Paxton2013}.  We chose
this modern stellar package, as, due to ease of use, it is becoming
widely popular, and may soon become  the most commonly used stellar
tool.  For the purpose of comparison, we clarify that, by default, we
use solar metallicity, and employ a standard set of stellar equations:
the simple grey atmosphere boundary condition with radiation pressure
at zero optical depth taken from \cite{cox_giuli}, mixing-length
theory (MLT) as in \cite{cox_giuli}, and radiative opacity tables
based on OPAL and \cite{ferguson05}.  If there is any deviation from
this default set, it will be clarified for each numerical experiment.
While some numerical issues that we will discuss in this paper are
relevant only to our choice of the stellar code, most of the issues
are relevant to modeling of fast ML  with any detailed stellar
code. We are aware that {\tt MESA} was  developed to evolve normal
stars, in a conventional way, and in this paper we will analyse how
{\tt MESA}, and possibly similar codes, work at an extreme beyond its
design. 

While the response of the donor's radius itself to ML is the first
important issue, the second foremost is how fast stellar material,
overfilling the  donor's Roche lobe, can be transferred to its
companion, and hence be effectively lost from the donor.  In \S3 and
\S4  we discuss treatment of the ML in detailed stellar models and
describe our adopted model of an optically thick stream appropriate to
a case of fast ML.  In \S5 we provide the results of our ML
simulations.  This includes a comparison with the previously published
results, and analysis of the stability of the MT in a binary system.

\section{Surface superadiabatic layer}

A major feature distinguishing adiabatic calculations of the MT
  from the results of detailed stellar codes for convective donors is
  the retention of a thin (in mass sense) layer of lowered entropy
  near the surface. Understanding the physics of this layer is likely to
  be key to understanding a realistic stellar response to ML.

\subsection{Definition} 

The term  ``superadiabatic'' makes sense only within the
  adopted convection theory, MLT.

In MLT, the relative efficiency of heat transfer via two mechanisms (radiative and convective)
can be characterized by the ratio of radiative and convective conductivites \cite[see equation 7.12 in][]{kw1994}:
\begin{equation}
\frac{\sigma_{\rm{rad}}}{\sigma_{\rm{conv}}} = \frac{3 a c T^{3}} {c_{P} {\rho}^{2}\kappa l^{2}_{\rm{m}}}
\sqrt{\frac{8H_{P}}{g\delta}},
\label{eq:conv_efficiency}
\end{equation}
where $l_{\rm{m}}$ is the mixing length and $H_{P}$ is pressure scale
height.  $\delta = - \left({\partial \ln \rho}/{\partial \ln
  T}\right)_{P}$ is a quantity that characterises the thermal
expansion properties of a medium.

In the framework of MLT, convective efficiency directly depends on two
free parameters -- on how far a blob rises or sinks before it mixes
with the environment (which can be seen in
Equation~\ref{eq:conv_efficiency} directly as $l_{\rm m}$), and on the
linear size of convective blobs (used in
Equation~\ref{eq:conv_efficiency}).  These free parameters are best
calibrated from observations and are not obtained from physical
considerations.
 
In convective zones overadiabaticity, which is
the difference between the actual and adiabatic temperature gradients,
$\nabla - \nabla_{\rm ad}$, is positive. We define the superadiabatic layer
as the region where overadiabacity is comparable to the gradients themselves.
 
\begin{figure}
\includegraphics[width=84mm]{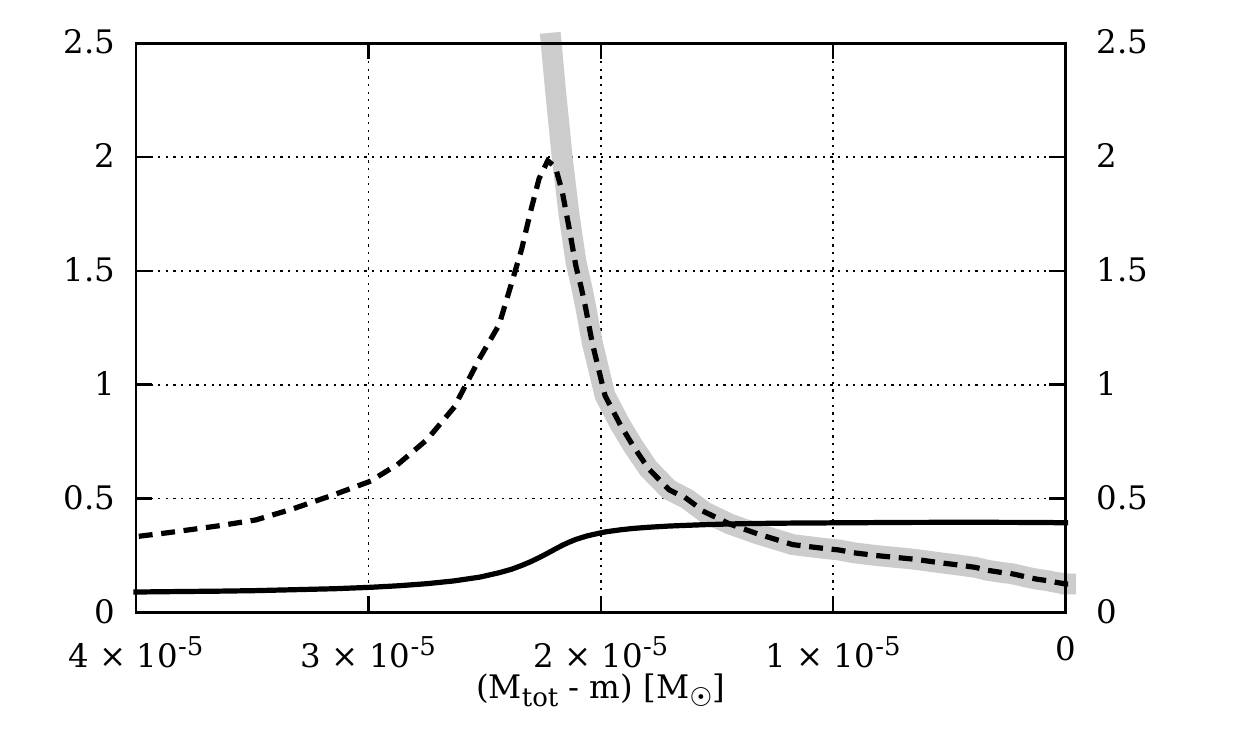}
\caption{Interplay of temperature gradients $\nabla$,
  $\nabla_{\rm{rad}}$ and $\nabla_{\rm{ad}}$ in the vicinity of the
  hydrogen ionisation boundary of a $70~\rsun$, $5~\msun$ red
  giant. The radiative gradient is shown with a solid grey line,
  adiabatic gradient -- solid black line, actual gradient -- dashed
  line; $m$ is the mass coordinate and $M_{\rm{tot}}$ is the total
  giant's mass.  Near the surface, where $\nabla_{\rm{rad}}<
    \nabla_{\rm{ad}}$, this giant has a surface radiative zone.  A
    superadiabatic zone is where $\nabla_{\rm rad}>\nabla_{\rm{ad}}$
    but convection is not efficient enough, and the actual gradient
    $\nabla$ exceeds $\nabla_{\rm{ad}}$ by a value comparable to the
    gradient itself. }
\label{fig:gradients}
\end{figure}

In this paper, for the efficiency of convection, we adopt the relation
between conductivities of convective and radiative heat transfer
denoted by $\sigma_{\rm{conv}} / \sigma_{\rm{rad}} = 1/U$.  Larger
values of $U$ correspond to less efficient convection and accordingly
to a larger over-adiabaticity.  In red giants, due to a combination of
factors $U$ increases in the vicinity of the stellar surface to the
extent that the real gradient of temperature becomes substantially
detached from the adiabatic gradient ($\nabla>\nabla_{\rm ad}$) (see
Figure~\ref{fig:gradients}).

\subsection{Superadiabatic layer and the effect of partial ionisation} 

\label{paper_sal}

\begin{figure}
\includegraphics[width=84mm]{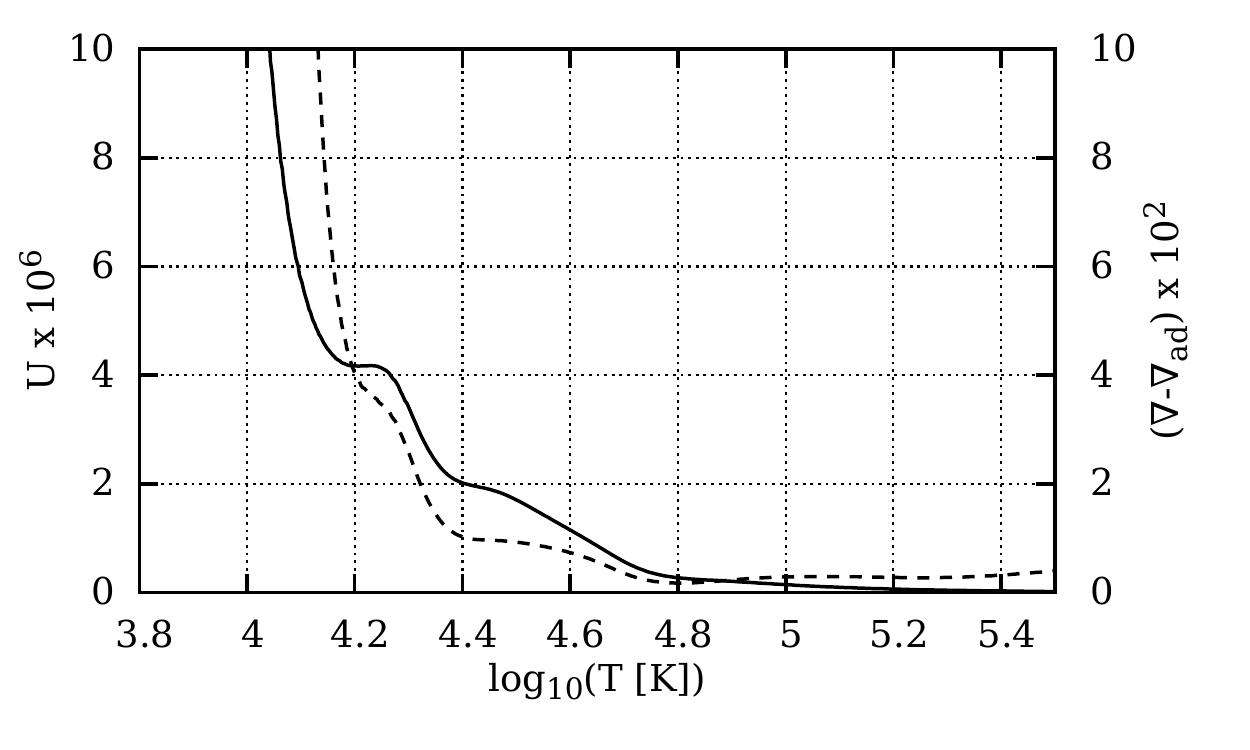}
\caption{Overadiabaticity $\nabla-\nabla_{\rm ad}$ (solid) and $U$ (dashed)  
in the outer layers of a $5~\msun$ and $50~\rsun$ red giant.}
\label{fig:sal_u_sa}
\end{figure}

\begin{figure}
\includegraphics[width=84mm]{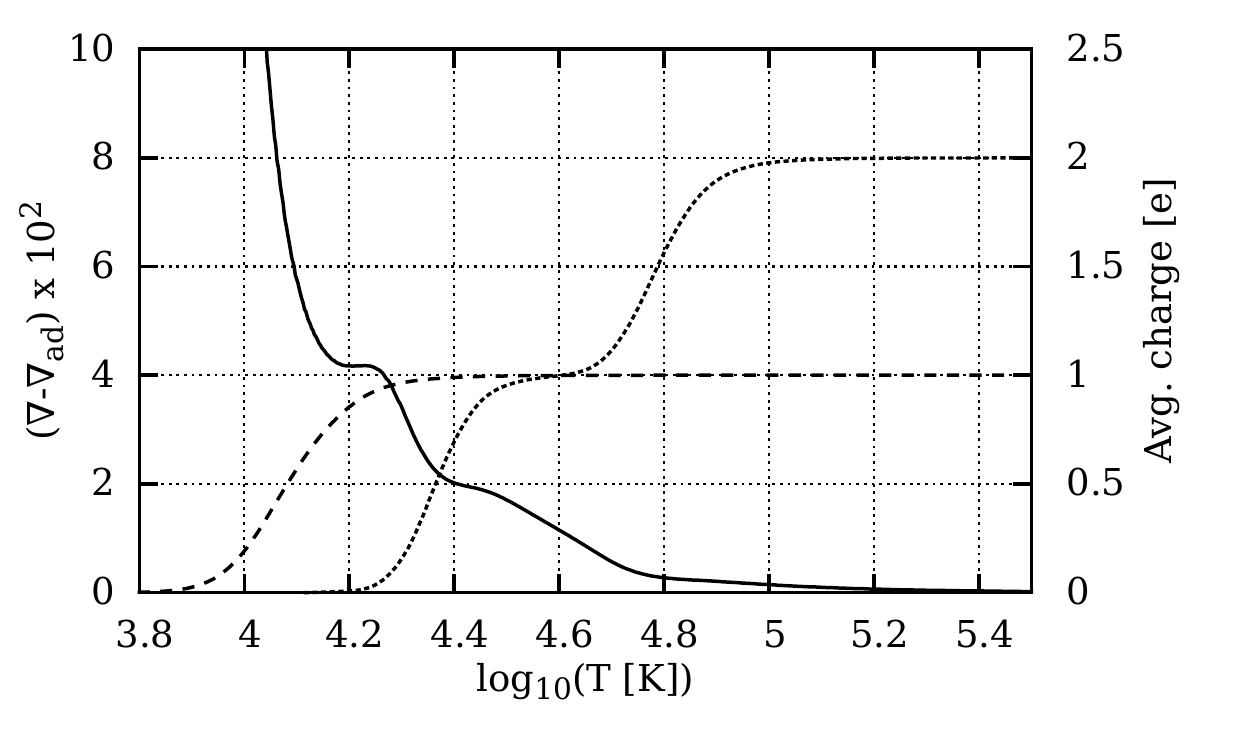}
\caption{Overadiabaticity $\nabla-\nabla_{\rm ad}$ (solid) and average charges of helium 
(dotted) and hydrogen (dashed) in the outer layers of a $5~\msun$ and $50~\rsun$ red giant.
Note that behaviour of overadiabaticity and partial ionisation  zones is strongly coupled.}
\label{fig:sal_io_sa}
\end{figure}

\begin{figure}
\includegraphics[width=84mm]{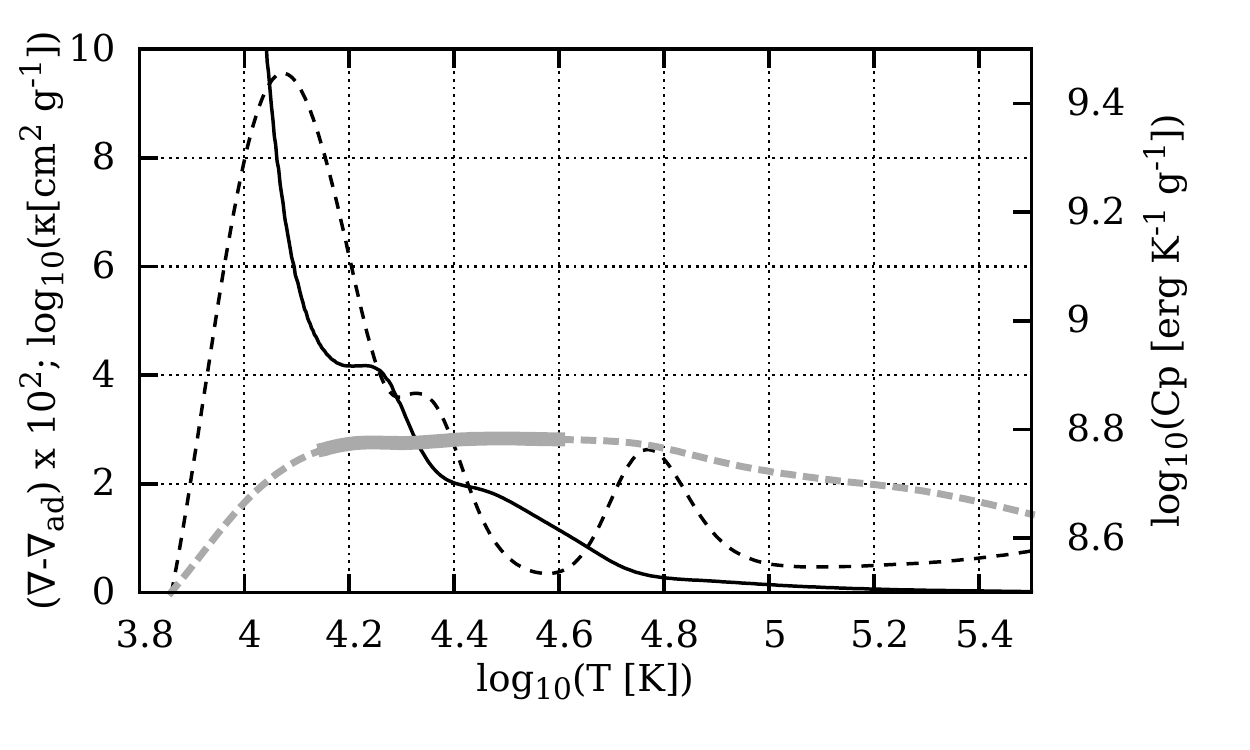}
\caption{Overadiabaticity $\nabla-\nabla_{\rm ad}$ (solid black), Rosseland mean opacity $\kappa$ 
(grey solid in the area of interest and grey dashed elsewhere)
and specific heat capacity at constant pressure $c_{P}$ (dashed black) in the outer layers of a 
$5~\msun$ and $50~\rsun$ red giant.}
\label{fig:sal_cp_sa_opac}
\end{figure}

Let us consider in detail what affects the behaviour of the efficiency
of the convection $1/U$ and of the overadiabaticity
$\nabla-\nabla_{\rm ad}$.  In the region where both quantities have
simultaneous slow-down in their almost monotonic decrease towards the
center (see Figure~\ref{fig:sal_u_sa}), hydrogen and helium change
their degree of ionisation (see Figure~\ref{fig:sal_io_sa}).  The
degrees of ionisation affect $U$ and $\nabla-\nabla_{\rm ad}$ by
strongly changing the heat capacity of the medium $c_{P}$ and the
Rosseland mean opacity (see Figure~\ref{fig:sal_cp_sa_opac}).

For the range of densities and temperatures in the envelope, the heat
capacity of a partially ionised gas is higher than that of both
neutral and fully ionised gases (ionisation consumes a part of the
heat transferred to the gas without increasing the kinetic energy of
its molecules).  The three major jumps in $c_{P}$, from highest to
lowest temperature, are associated with the following recombinations:
\begin{enumerate}[(i)]
\item double ionised helium $\rightarrow$ single ionised helium;
\item single ionised helium $\rightarrow$ neutral helium;
\item ionised hydrogen $\rightarrow$ neutral hydrogen.
\end{enumerate} 


Let us discuss the reason for two overadiabaticity plateaus in the
area of interest marked with the solid opacity curve in
Figure~\ref{fig:sal_cp_sa_opac}.  In this area, partial recombination
of elements affects the convective efficiency mainly through $c_{P}$,
because opacity does not vary much there.  Due to the change in
monotonicity of $c_{P}$ at half-ionisation, recombination of either
hydrogen or helium impedes the growth of overadiabaticity towards the
surface while the ionisation fraction is between $100\%$ and $50\%$,
but when the ionisation fraction is lower than $50\%$ the
recombination, on the other hand, accelerates the growth of
overadiabaticity (see Equation~\ref{eq:conv_efficiency} and
Figure~\ref{fig:sal_cp_sa_opac}).  Because of this, the areas where
the growth of overadiabaticity is impeded (plateaus) are shifted with
respect to the maxima of $c_{P}$.  The initial stage of partial
recombination of hydrogen is able to completely suppress the growth of
overadiabaticity towards stellar surface in the envelope; the initial
recombination of helium also impedes it to a large extent.  The shape
of the superadiabatic layer, hence, is substantially affected by the
recombination of hydrogen and helium.

At  the final  stages  of hydrogen  recombination,  the opacity  falls
sharply by  a few  orders of  magnitude.  As  a result,  the radiative
gradient, which is proportional to  opacity, falls below the adiabatic
gradient  (Figure~\ref{fig:gradients}).   We   call  this  region  the
"surface radiative zone". It takes place outside the temperature range
shown in Figures ~\ref{fig:sal_u_sa}-\ref{fig:sal_cp_sa_opac}, at $\lg
T < 3.73$.  The  spatial size of the surface radiative  zone is of the
order of  the mixing  length.  Hence it  is plausible  that convective
blobs overshoot the boundary between  the superadiabatic layer and the
surface radiative zone \citep[see also][]{kuhfuss86}.

\subsection{Mass of superadiabatic layers in convective donors}

\begin{figure}
\includegraphics[width=84mm]{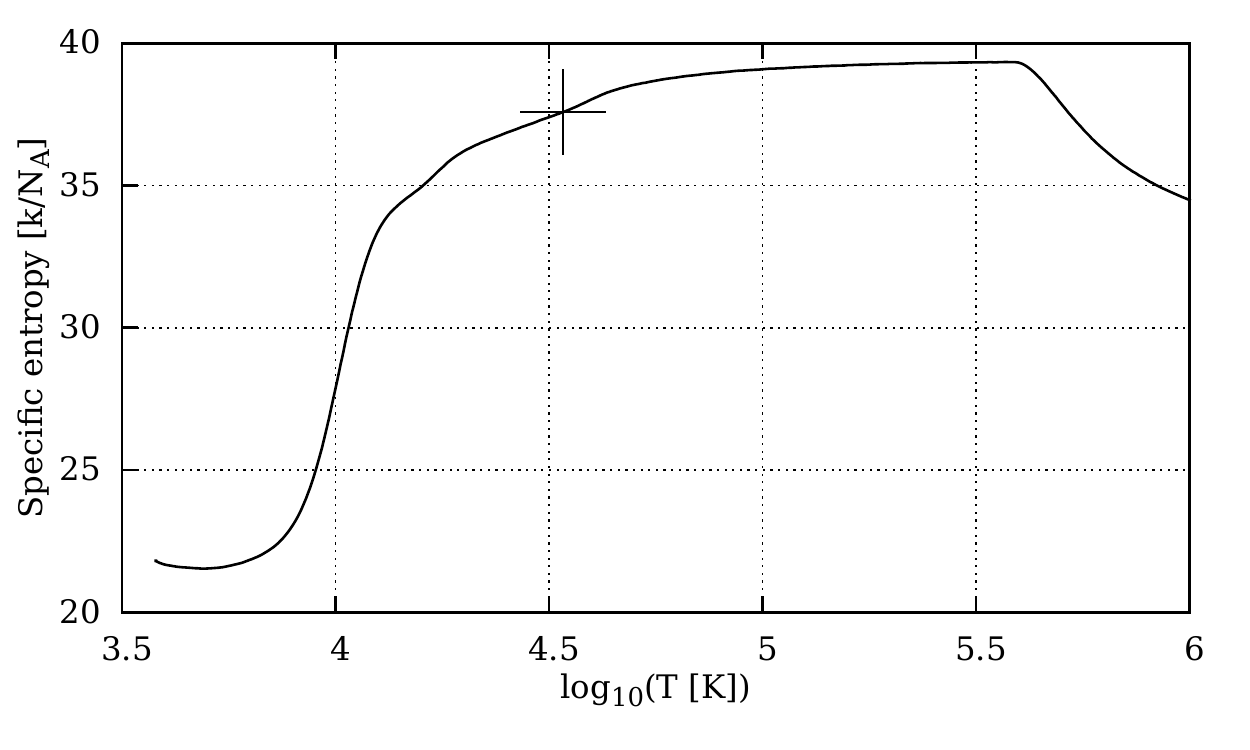}
\caption{Specific entropy in the outer layers of a $14.6~\msun$ and $500~\rsun$ red giant, the cross denotes the bottom of the superadiabatic layer
calculated as per Equation~\ref{eq:bottoms}.}
\label{fig:sal_ent}
\end{figure}

For  analysis  of  superadiabatic   layers  and  radiative  zones,  we
introduce the  quantity $m_{\rm sad}$,  the mass of these  layers.  We
measure $m_{\rm sad}$  from the stellar surface to the  point near the
start of the entropy drop, $m_0$.  Since the convection is never fully
adiabatic, the entropy profile in the  envelope is not flat, and there
is no  sharp transition  between the entropy plateau and  the entropy
drop (see  Figure~\ref{fig:sal_ent}).  For integrity between  stars of
different masses,  we therefore choose to  define $m_0 $ as  the point
where entropy is

\begin{equation}
\label{eq:bottoms}
S(m_{0}) = S_{\rm min} + 0.9(S_{\rm conv} - S_{\rm min}).
\end{equation}

\noindent Here $S_{\rm min}$ is the local {\it minimum} of
entropy in the outer layer of the star and $S_{\rm conv}$ is the (maximum) value of entropy in
the convective envelope.

\begin{figure}
\includegraphics[width=84mm]{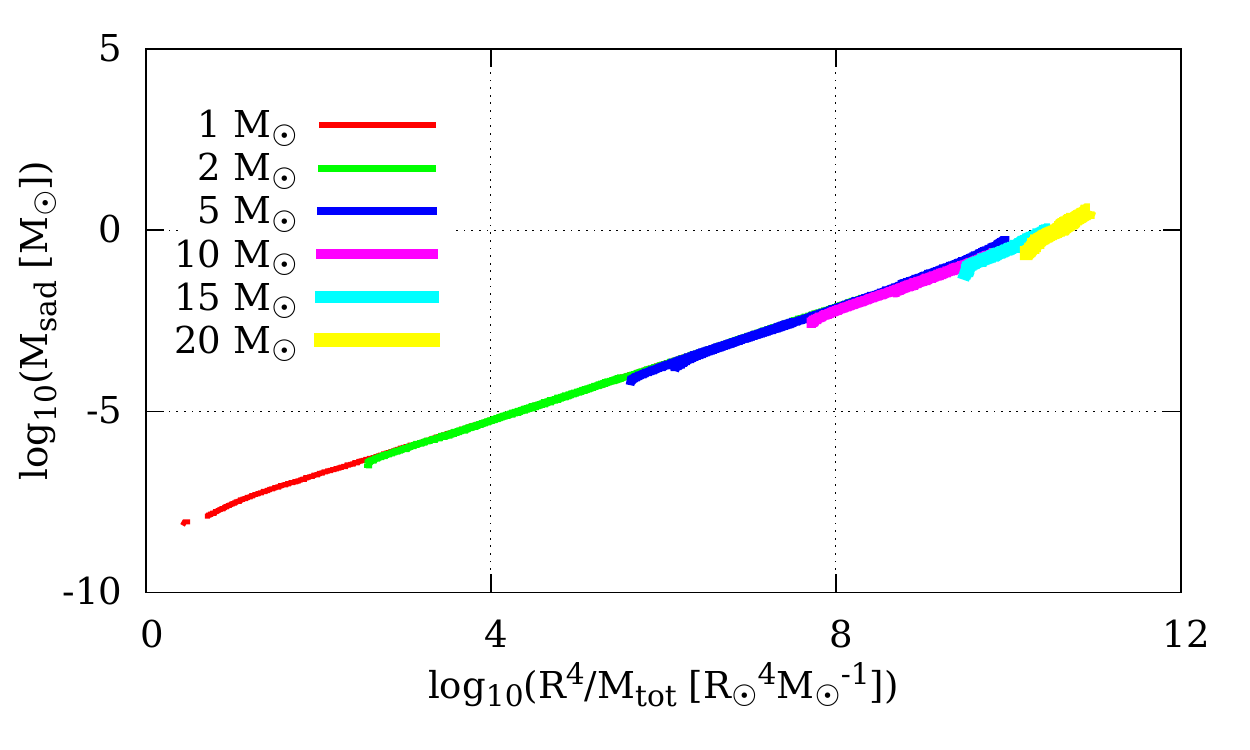}
\caption{Total mass of superadiabatic layers and surface radiative zones in stars with 
convective envelopes of different masses and ages.}
\label{fig:sad_mass}
\end{figure}

Clearly, it can be seen  in Figure~\ref{fig:sad_mass} that the mass of
this shell varies between giants  of different masses and radii.  From
the  behaviour   of  the  envelope's  simplified   analytical  solution
\citep[see   \eg    Figure~10.2   in][]{kw1994},   we    expect   that
overadiabaticity of convective envelopes  becomes substantial at about
the  same pressure,  to an  order of  magnitude. Indeed,  in realistic
stellar  models for  stars of  different masses,  the pressure  at the
bottom of the  superadiabatic layer does not vary  much from $P(m_{0})
\approx 10^5 $dyn/cm$^2$.  In this case, we can estimate $m_{\rm sad}$
by simply  considering a pressure  drop.  The equation  of hydrostatic
equilibrium over the superadiabatic layer and radiative zone is

\begin{equation}
\frac{dP}{dm_{\rm sad}} = \frac{G(m_0 + m_{\rm sad}(r))}{4 \pi r^4}.
\end{equation}
As the pressure at the top of the superadiabatic layer (photospheric pressure) is much 
less than at its bottom, we find 
\begin{equation}
\label{eq:sad_mass}
m_{\rm sad}(R) \approx \frac{4 \pi R^4 P_0}{GM_{\rm tot}} \approx 1.1 \times 10^{-10} M_\odot \, \frac{ (R/R_\odot)^4}{(M_{\rm tot}/M_\odot)}.
\end{equation}

\noindent Indeed,  as illustrated in  Figure~\ref{fig:sad_mass}, stars
with convective envelopes at {\it any} evolutionary stage have 
$m_{\rm  sad}$ which  can be found  using this relation.   This approximation
has  only  limited, though  still  very  good, applicability  to  more
massive stars, where $m_{\rm sad}$ is  comparable to the total mass of
the  star's envelope  (we note  also  that it  is well  known that  in
massive  stars   almost  the  entire  convective   envelope  could  be
superadiabatic).

It is  important to note that  this is the mass  of the superadiabatic
layer and surface radiative zone in a star {\it unperturbed} by ML.  Mass loss
affects both our  assumptions about the pressure drop (as  the star is
not in hydrostatic equilibrium anymore), and the entropy profile.

We see from the preceding analysis that while the mass of the
  superadiabatic layer is relatively small in low-mass giants, it grows
   for more massive stars, increasing substantially as
  the star is evolving and expanding. Stellar models that have a
  constant entropy profile are therefore not justified, as the error
  in the case of massive stars could be enormous.  We will also show in
  \S\ref{paper_massloss_eps_g} that the mass of the superadiabatic
  layer helps estimate its rate of energy generation for each
  ML rate, and how to find the timesteps that provide valid
  results for the radial response on the ML.

\subsection{Energetics of the superadiabatic layer} 
\label{paper_massloss_eps_g}

We saw a hint in \S~\ref{paper_sal} that the formation of a
  superadiabatic layer is to a certain extent governed by
  recombination, hence this layer might be
  energetically important for obtaining the correct ML
  response. Energetically, recombination enters the structure
  equations through the gravitational energy term, which is the last
  term in the energy generation equation and is defined as:

\begin{equation}
\epsilon_{\rm{g}} = -T \left (\frac{\partial s}{\partial t}\right )_m
\label{eq:eps_g2}
\end{equation}

The  $\epsilon_{\rm{g}}$ term can be found directly from the
  Equation \ref{eq:eps_g2},  using only one temporal derivative  of
  the entropy which is derived from the EOS.  Alternatively,
  $\epsilon_{\rm{g}}$ can be calculated by subtracting two terms that
  depend on  temporal derivatives of thermodynamic quantities, for
  example:

\begin{equation}
\epsilon_{\rm{g}} = -T c_{P} \left[(1 - \nabla_{\rm{ad}} \chi_{T}) \frac{1}{T}\left(\frac{\partial T}{\partial t}\right)_m -
\nabla_{\rm{ad}} \chi_{\rho} \frac{1}{\rho} \left(\frac{\partial \rho}{\partial t}\right)_{m}\right]
\label{eq:eps_g1}
\end{equation}

When mass is removed from the surface of a red giant, the partial ionisation zones of
 hydrogen and helium discussed in \S~\ref{paper_sal} move to areas that were previously fully ionized, deeper in the star. Recombination of these previously
 ionised areas leads to the release of recombination energy.
One can estimate that the recombination energy of pure hydrogen is  
$W \sim 10^{13}$~erg~g$^{-1}$. 
For the ML rate of ${\dot M } = 10^{-2}$~$M_\odot$~yr$^{-1}$ the 
recombination energy release should be of the order of $W {\dot M} \sim 6\times 10^{36}$~erg~s$^{-1}$. 

\begin{figure}
\includegraphics[width=84mm]{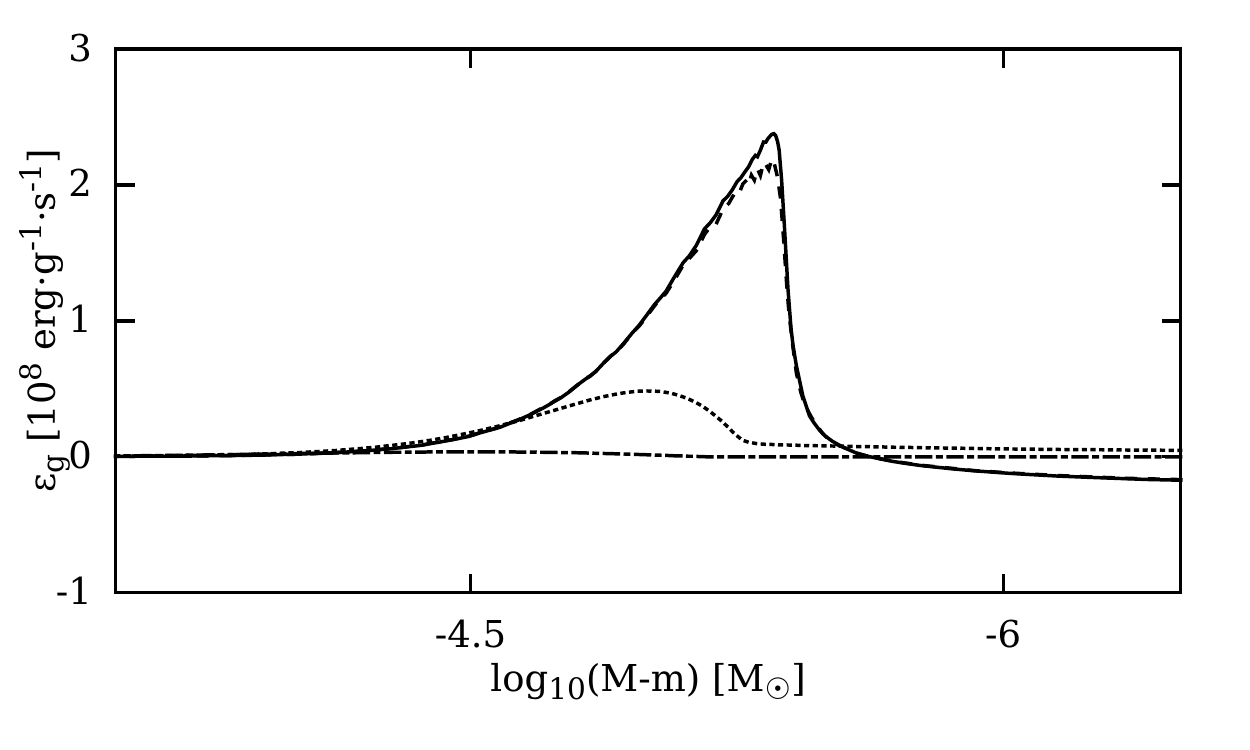}
\caption{Values of $\epsilon_{\rm{g}}$ in the same $5.0~\msun$,
  $50~\rsun$ red giant subjected to the same constant ML rate of
  $10^{-2}~\msun$~yr$^{-1}$ shown at the same moment of time:
  $0.1~yr$ since the start of ML. Dash-dotted, dotted, dashed and
  solid lines correspond to constant timesteps $10^{-1}$, $10^{-3}$,
  $10^{-5}~yr$ and $10^{-6}~yr$ respectively. The observed
    discrepancy in the values of $\epsilon_{\rm{g}}$ is due to the
    inaccuracy of the first order formula used to
    calculate Lagrangian derivatives using the Equation~\ref{eq:eps_g2}
    (see Section~\ref{paper_massloss_eps_g} for details).
}
\label{fig:eps_g}
\end{figure}

Let us consider now this mass loss on an example of a $5.0~\rm{\msun}$ and $50~\rm{\rsun}$ red giant.
Using our Equation~\ref{eq:sad_mass} for the mass of the superadiabatic layer, we find that 
its total mass is about $10^{-4}~\rm{\msun}$, though the part where hydrogen 
recombination energy is released is usually only a fraction of it,
 about $10^{-5}~\rm{\msun}$ (see also Figure \ref{fig:eps_g}). 
The contribution of this recombination component alone in $\epsilon_{\rm g}$ in
such a superadiabatic layer is expected to be of the order of  $3\times 10^{8}$~erg~g$^{-1}$~s$^{-1}$. 

We have checked what values of $\epsilon_{\rm{g}}$ we obtain {\it
    in practice},  using the detailed stellar code of our choice, {\tt
    MESA}.  First, we tested the method where the entropy derivative
  is found using  Equation~\ref{eq:eps_g2}, as it takes into account
  composition changes, unlike the second method that uses
  Equation~\ref{eq:eps_g1}. However, this method suffers
  from an inexact calculation of entropy from the EOS. After
  performing numerical tests, we found it to not be suitable for the
  small timesteps necessary for fast mass loss calculations:  as the
  timesteps become smaller, the errors in the entropy values become
  comparable  to their Lagrangian differences between subsequent
  timesteps. The two-component formula~\ref{eq:eps_g1} always uses structural variables.
  Entropy, used in the formula~\ref{eq:eps_g2}, is never a structural variable: it is always derived
  from the structural variables through the tabulated EOS. In addition, composition artefacts
  can play a role in entropy fluctuations.

Second, we tested the method where the entropy derivative is found
using Equation~\ref{eq:eps_g1}, see Figure~\ref{fig:eps_g}.  We found
that the expected physically reasonable level of $\epsilon_{\rm g}$ is
only reached when the timestep is below $\approx 10^{-3}$~yr.  We
relate it to the inaccuracy of the first-order numerical differentiation formulae 
used in practice to calculate the right-hand side of
(\ref{eq:eps_g1}) at larger timesteps. Note that the right-hand side of
(\ref{eq:eps_g1}) is effectively proportional to the Lagrangian derivative of entropy\footnote{If one
  neglects composition changes.}, which is calculated indirectly
through the corresponding derivatives of $T$ and $\rho$.  Therefore,
the error in calculation of $\epsilon_{\rm g}$ with formula
(\ref{eq:eps_g1}) is, as expected, the most significant in the
superadiabatic layer and surface radiative zone,  because the second
Lagrangian derivative of entropy is the highest there\footnote{ We'd
  also like to mention that quite recently a new scheme for the
  calculation of  $\epsilon_{\rm g}$ was implemented in {\tt MESA},
  that follows a mixed Lagrangian-Eulerian approach. We performed a
  preliminary testing of this scheme on mass-losing red giants.
  Unfortunately, this new scheme is not suitable for our simulations
  yet because of  severe numerical artefacts that it introduces under
  certain conditions.  These artefacts have to be carefully studied
  and eliminated before we can consider  incorporating this new scheme
  into our simulations.  Because of this, we only discuss the standard
  purely Lagrangian scheme for the calculation of $\epsilon_{\rm g}$
  in this paper.}.

We do not suggest a complete formal procedure to calculate the errors
in calculation of $\epsilon_{\rm{g}}$.  Instead, we resort to a plain
comparison of results obtained with various timestep selection
approaches.  For example, the comparison for one of our models
(Figure~\ref{fig:eps_g}) shows  that if it loses not more than
$\approx1/1000$ of its initial $m_{\rm sad}$ in one timestep,  then
the calculation of $\epsilon_{\rm{g}}$ in the superadiabatic layer is
affected only marginally.  Removing mass in such small timesteps is,
however, quite resource intensive.  We can foresee that the outlined
numerical problem with a recombination zone that moves too fast in
mass can be eventually resolved using a technique similar to the one
used for calculating thin shell burning as in 
\cite{Eggleton67,Eggleton73}.  However, for the purpose of the studies
  presented in this paper, we can afford to use a numerically
  intensive way.

   In Section~\ref{paper_sim_results_const}  we will show  how
  inaccurate calculations of $\epsilon_{\rm{g}}$ affect  the radial
  response to the mass loss.   While the  numerical problems that  we
  described  in this section are native only to the code we use, when
  fast mass loss rate calculations  are performed  with  another
  code, the numerically  obtained $\epsilon_{\rm{g}}$ has to be tested
  against its expected value that can be found using the method
  described in this subsection.

 \section{Fast ML in an one-dimensional star}

There currently does not exist a comprehensive and
  self-consistent hydrodynamical simulation to treat the
  three-dimensional problem of MT in a binary. 
  Even when three-dimensional simulations eventually become self-consistent, they
  would likely resolve only the question of the initial stability of
  the Roche lobe overflow (RLOF), but not the long-term MT, which will remain a
  prerogative of one-dimensional stellar codes in the foreseeable
  future.

  In a standard one-dimensional ML model, the stellar codes use the
  regular set of structure equations, but adopting the boundary
  condition that the total mass decreases with time.  This boundary
  condition is an unavoidable reduction of the three-dimensional
  picture of the ML to one dimension.  In other words, $\dot M$
  represents our best understanding of the stream that is formed in
  the vicinity of $L_1$ and that carries the donor's material away from
  the donor.  In this Section we examine the reaction of giants to the
  ML in one-dimensional stellar codes, explaining in particular
  the nature of the feature observed in the previous MT calculations.

\subsection{Understanding the initial contraction of a red giant upon the instantaneous ML in 1D stellar codes} 

\label{paper_massloss_hresponse}

Two completely opposite responses to ML were found in detailed 1D simulations by \cite{Passy2012}.
According to them, hydrostatic stellar models expand in response to ML,
while models with hydrodynamical terms, on the opposite, shrink.

In neither of these two approaches the ML experiment is close
to what would happen in Nature, where MT never starts abruptly at some
fixed high MT rate. However, it is important to understand what causes
the dramatic difference between these two approaches.

 \cite{Passy2012} provided the
  following explanation: \textit{"...some energy that is stored in
    gravitational form in the hydrostatic models is actually in a
    kinetic form [in hydrodynamic models], leading to the star
    contracting instead of expanding"}, although how exactly the
  transformation of gravitational energy into kinetic leads to contraction
  was not explained. 

Instead of the energy argument, we argue that the main reason  
for the radial response is due to the material being consumed from the 
surface of a red giant with linear velocity ${\dot m} / (4 \pi r^2 \rho)$, 
where $r$ is the radius of a red giant, and $\rho$ is
the surface density. Note that this term intrinsically decreases the radius.

\begin{figure}
\includegraphics[width=84mm]{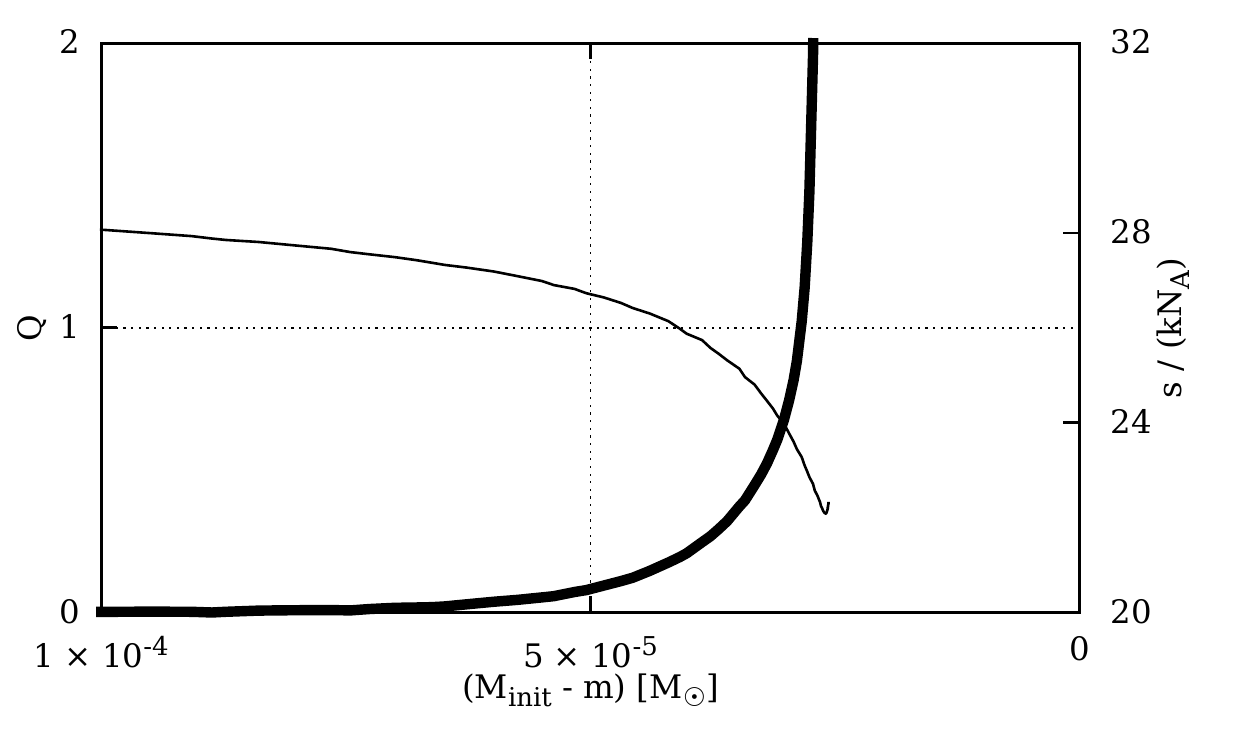}
\caption{A red giant of initial mass 5.0~$\msun$ and radius 50~$\rsun$ 
subjected to ML of 0.1~$\msun$~yr$^{-1}$. The entropy profile is shown with a thin line,
the ratio $Q$ of acceleration and gravitational terms is shown with a thick line. 
It is clear that  $Q$ is far from being negligible compared to unity within the superadiabatic
layer and surface radiative zone.}
\label{fig:hrad}
\end{figure}

To validate that this term is dominant, we need to consider
  the involved stellar equations.
A standard way to introduce hydrodynamical treatment into a stellar
code is to use a truncated Navier-Stokes equation in spherical
coordinates to calculate the hydrodynamical term of the pressure
derivative, for example:
\begin{equation}
\left[\frac{\partial P}{\partial m}\right]_t = -\frac{Gm}{4 \pi r^4} \left(1 + Q \right),
\label{eq:sem}
\end{equation}
where the acceleration term $Q$ is the ratio of the local Lagrangian 
acceleration to the local gravitational acceleration:
\begin{equation}
Q = \left[\frac{\partial^2 r}{\partial t^2}\right]_m / \frac{Gm}{r^2} = \frac{a(m, t)}{GM / r^2}, 
\end{equation}
where $a(m, t)$ is the local Lagrangian acceleration.
For a star in hydrostatic equilibrium, $Q = 0$.
In Figure~\ref{fig:hrad} we illustrate how significant the acceleration 
term can be in the superadiabatic layer and surface radiative zone of a red giant at high ML rates.

Let's now consider the initial response after some (small compared to the dynamical timescale) 
time $\tau$ to ML ${\dot m} < 0$ of a red giant
of initial mass $M_{0}$ without artificial viscosity or any other effects that would alter Equation~(\ref{eq:sem}).

If $P(m, t)$ is a smooth and bounded function and $m(t)$ and $r(t)$ are continuous and bounded, then
\begin{equation}
\label{eq:accel_lim}
\lim_{t \to 0} a(m,t) = 4 \pi R_{0}^2  \left[\frac{\partial P}{\partial m}\right]_{t = 0} + \frac{GM_{0}}{R_{0}^2} 
\end{equation}
By integrating Equation~(\ref{eq:accel_lim}), we get for the velocity $v$ at the surface\footnote{Here we use the "little-o" notation for the vicinity of zero. A brief definition of this notation is $f(t) = o(g(t)) \Leftrightarrow \lim_{t \to 0} f(t) / g(t) = 0$ \cite[for details see, e.g.,][]{Kevorkian85}.}:
\begin{align}
\label{eq:v_exp}
v(t) =  v_{\rm 0} +  \left( 4 \pi R_{0}^2 \left[\frac{\partial P}{\partial m}\right]_{t = 0} + \frac{GM_{0}}{R_{0}^2} \right) t \\ \nonumber
+ o(t),
\end{align}
where $v_{0} = v(0)$.
To obtain the complete radial response, this velocity must be combined with the material consumption velocity, which is equal to
\begin{equation}
v_{\rm{c}}(t) = \frac{\dot{m}}{4\pi\rho_{0}R_{0}^2} + o(t),
\end{equation}
where $\rho_{0}$ is the initial surface density of the star. Their sum must in turn be integrated in time.
Thus, the complete radial response of a red giant to ML is given by:
\begin{eqnarray} 
\label{eq:r_exp}
R(t) &=& R_{0} + \left( \frac{{\dot m}} {4 \pi R_{0}^2 \rho_{0} } + v_{\rm 0} \right) t 
\\ \nonumber   &+&  \left( 2 \pi R_{0}^2 \left[\frac{\partial P}{\partial m}\right]_{t = 0} + \frac{GM_{0}}{2 R_{0}^2} \right)t^2  + o (t^2)
\end{eqnarray}

If the original star was in hydrostatic equilibrium, then $v_{0} = 0$, and, according to Equation~(\ref{eq:accel_lim}),
the term,  proportional to $t^2$ in Equation~(\ref{eq:r_exp}), 
would be equal to $o(1)t^2 = o(t^2)$, and hence 
is vanishing.  The total radial response  reduces to

\begin{equation}
R(t)=R_{0} + \frac{{\dot m}} {4 \pi R_{0}^2 \rho_{0} } t +  o (t)
\label{eq:consump_der_red}
\end{equation}

This result may be directly compared to the decrease in the radius of
a stellar model that occurs during the first timestep after the start
of the ML, divided by this timestep.  This comparison helps to
  understand the role of the adopted outer boundary condition (see
  Sections~\ref{sec:acc} and \ref{paper_sim_results_const}).  
We conclude that the contraction observed in the 
  simulations of \cite{Passy2012}  is neither a numerical
  error, nor a conversion of gravitational energy to kinetic, but a
  natural consequence of a finite pace of expansion of the
  one-dimensional envelope that was forced to experience a fast ML.

\subsection{The acceleration term}

\label{sec:acc}

As shown in Figure~\ref{fig:hrad}, the acceleration term is
  becoming dangerously large in the surface layers during fast ML. In this
  Section we describe the modification of stellar equations and the
  boundary condition to take this term into account.

To obtain the pressure at the outer boundary $P_{\rm{bc}}$ in a simple
grey  atmosphere approximation,  it  is common  to  use the  following
well-known  formula\footnote{Here  for  simplicity we  omit  radiation
  pressure  at zero  optical  depth.  This term  may  be important  if
  radiative  pressure   is  a   substantial  fraction  of   the  total
  pressure.}:

\begin{equation}
P_{\rm bc} = \frac{2}{3} \frac{GM}{R^2 \kappa}
\label{eq:grey_orig}
\end{equation}

Unfortunately, the above formula is based on the equation of hydrostatic 
equilibrium, hence in our case it is not applicable,
not to speak that it is inconsistent with the Equation~\ref{eq:sem} 
that is used in many stellar codes.
For this reason we took into account the acceleration term not only 
in the equation of motion~(\ref{eq:sem}), but also to find the
boundary condition for a simple grey atmosphere, which we use in our 
experimental version of {\tt MESA}:
\begin{equation}
P_{\rm bc} = \frac{2}{3} \frac{GM}{R^2 \kappa} \left( 1 + Q \right)
\label{eq:grey_corr}
\end{equation}

\noindent Numerical experiments with {\tt MESA} showed that the
  choice of the surface boundary condition affects the resulting
  radial response. The modified boundary
  condition must be used because in this case we obtain a radial response
  substantially closer to that predicted by 
  Equation~\ref{eq:consump_der_red} (see Section~\ref{paper_sim_results_const}).

Similarly, we take into account the acceleration term for the temperature equation 
\begin{align}
\left[\frac{\partial T}{\partial m}\right]_t = -\frac{G m}{4\pi r^4}\left(1 + Q \right)  \frac{T}{P} \nabla,
\end{align}
for convective conductivity in the Mixing Length Theory, and for the radiative gradient:
\begin{equation}
\nabla_{\rm rad} = \frac{3 \kappa PL}{16 \pi a c T^4}  \frac{1}{GM \left( 1 + Q \right) }\ .
\end{equation}

\section{Rate of the mass transfer} \label{paper_massloss_scml}

We describe below an optically thick model for MT that we have adopted
for our studies of  hydrodynamic response to rapid MT in binaries.

We follow a  conventional way to calculate the MT  rate by integrating
the mass flow over a ``nozzle'' cross-section that is taken on a plane
perpendicular to the line connecting the  centres of the two stars and
passing through the $L_1$ point:

\begin{equation}
{\dot   m}  =   \int_{\rm   nozzle}  v_{\rm   flow}  \rho_{\rm   flow}
 dS \end{equation}

\noindent Here $v_{\rm flow}$ is the velocity that the stream has within 
the nozzle, and  $\rho_{\rm flow}$ is the density that  the stream has
at  the  same   position.   We  note  that  the  use   of  the  nozzle
cross-section only  in the $L_1$ neighbourhood  implies that only the $L_1$ MT
rate can be found,  and the rate of ML via  $L_2/L_3$ overflow cannot
be calculated.

Furthermore,  any  scheme that  finds  an  ML boundary  condition  for
stellar one-dimensional  codes adopts  some set of  simplifications to
find  the distribution of the stream's density  and velocity  throughout the
nozzle, as  well as the nozzle  geometry.  The flow is considered  to be
steady,  which permits  the  use  of the  Bernoulli  theorem along  the
streamlines. The  standard assumptions  are that  the nozzle  at $L_1$
coincides      with      the       sonic      surface      of      the
flow \citep{1976ApJ...207L..53L}, and that  initial velocities are negligible 
at the origin of the streamlines. We adopt the same assumptions.

The MT rate then becomes dependent only on these assumptions: 

\begin{enumerate}[(i)]
\item the adopted {\it geometry} of the donor and the nozzle; 
\item the {\it streamlines}   --  this assumption, coupled  with the adopted evolution of the specific entropy
along the streamlines, allows relating the donor's thermodynamical  
properties to those of the flow  crossing the nozzle.
\end{enumerate}

\subsection{RLOF formalism and geometry of the problem}

The fundamental  simplifying assumption  that governs the  whole Roche
lobe formalism is \emph{volume  correspondence}.  More precisely, this
is the assumption that  thermodynamical parameters (pressure, density,
composition,  etc.)  at  a  certain radius  $r$  in a  one-dimensional
stellar  model are  the same  in  three-dimensional space  at a  Roche
equipotential  whose  enclosed  volume is  $(4/3)\pi  r^3$.   Whenever
one-dimensional  stellar  evolution is  considered  in  terms of  RLOF
formalism, the volume correspondence assumption is automatically used.
Note  that when  a donor  experiences a  substantial RLOF  this volume
correspondence for thermodynamical parameters  can be applied only far
from the $L_{1}$ point, where the  donor's material is almost at rest.
The Roche  lobe volume radius $R_{\rm L}$ is usually found  using one of two
well-known approximations  \citep{1971ARA&A...9..183P,1983ApJ...268..368E}.   The
area of  the nozzle  is then  usually approximated by  the area  of an
ellipsoid  by taking  the  second-order term  in  the Roche  potential
expansion within the sonic surface near the $L_{1}$ point.

An additional  simplifying assumption is applied  when integrating the
ML  rate over  the potential region between  the Roche  lobe potential
$\Phi_{L}$  and the  photospheric  Roche  potential $\Phi_{\rm  phot}$
(note, that  $\Phi_{\rm phot} >  \Phi_{L}$).  At this point,  to avoid
volume integrations, it  is a common approach to  implicitly break the
volume    correspondence    assumption    and    replace    it    with
the \emph{pressure correspondence} assumption.

The   pressure  correspondence assumption  asserts   that  the   thermodynamical
parameters  (temperature, density  and composition)  at a  point A  in
three-dimensional  space  are  the  same  as  at  a  point  B  in  the
one-dimensional model,  provided that  the pressure at  the point  A in
three-dimensional space is the same as  the pressure at point B in the
one-dimensional model.   If one  applies this  pressure correspondence
far from the $L_{1}$ point and assumes, in addition, that

\begin{equation}
 \textbf{a} \cdot \grad{\Phi} \ll \left| \grad{\Phi} \right|^2,
\label{eq:cond_kolb2}
\end{equation}

\noindent where $\textbf{a}$ is the local Lagrangian acceleration, 
and $\Phi$ is  the local Roche potential, then it  becomes possible to
replace the integral over radius with the integral over pressure.  The
benefit  of the  pressure correspondence  assumption is  that one  can
avoid the painful calculation  of the differential $d(\Phi(r))$, which
requires volume  integrations.  We note that  once a star is  not in 
hydrostatic    equilibrium,   $Q\ll    1$   does    not   hold    (see
Section~\ref{paper_massloss_hresponse}),                           and
condition  \ref{eq:cond_kolb2} cannot  be satisfied.   Hence, pressure
correspondence should not be used for  rapid mass loss.

RLOF during  the evolution of a  mass-losing red giant can  however be
very substantial.  When an equipotential surface is far from the Roche
lobe surface, the expansions obtained in the vicinity of the $L_{1}$ point break down
since the expansions are truncated after the quadratic term. Higher-order terms are
no longer negligible.
Instead, we conduct the realistic Roche
lobe integrations which employ the Runge-Kutta-Fehlberg integrator and
the   damped  Newton-Raphson   solver   to   obtain  all   geometrical
parameters.  These integrations  have been  conducted for  275001 mass
ratios from  0.06 to  19.145 with an increment of  $7\cdot10^{-5}$.  For
each mass ratio we calculate the nozzle areas and volume radii for 200
equipotentials   lying  between   the   $L_{1}$   potential  and   the
$L_{2}/L_{3}$  potential. 
In  addition to  the use  of a  more precise
relation  between the star's volume  and  the Roche  lobe volume,  volume
correspondence  that   is  necessary  for  rapid  mass   loss  and a 
non-simplified nozzle shape, this also  allows us to track whether the
donor overfills  the $L_{\rm{2}}/L_{\rm{3}}$ equipotential  during the
mass transfer.

\subsection{Streamlines}

For a rapid mass transfer  rate, an "optically thick" approximation is
usually  used  \citep[see][and many  others]{Paczynski1972,  Savonije1978,
KolbRitter1990, Ge10}.  The  flow of matter towards  $L_{1}$ in this
approximation is adiabatic and  streamlines go along the equipotential
surfaces of  the Roche  potential.  The  photosphere corresponds  to a
Roche equipotential which  lies outside the Roche lobe.   Close to the
$L_{1}$ point,  the photosphere turns  into the outer boundary  of the
optically thick  flow which flows  across the sonic surface  along the
Roche  equipotential  surfaces  with  the  local  adiabatic  speed  of
sound \citep{KolbRitter1990}.

To find the stream's density and sonic  velocity at $L_1$, one is required
to adopt the evolution of the specific entropy in the flow along the streamlines. Often
the flow  is taken to  be polytropic -- that  is, the flow  preserves a
constant value of $P/\rho^{\gamma}$ along a streamline, where $\gamma$
is  the   adiabatic  exponent.   In  certain   cases,  the  polytropic
stratification of  the donor  itself is  adopted.  Note  that the stream's
velocity, while locally sonic, is not constant across the nozzle.  The
flow  also  does  not need  to  be  isentropic  if  the donor  is  not
isentropic  --  the streamlines  can  originate  at different  initial
equipotentials.

We also assume that a flow is  adiabatic.  An adiabatic flow is not, of
course, entirely polytropic due to,  for example, recombination,
that occurs as gas  flows towards the sonic surface.  It
means  that $\gamma$  varies along  an adiabat.   For this  reason, we
employ a realistic equation of state taken from {\tt MESA}, that among
other effects takes into account the ionisation of elements in the mix
and the radiative  component of pressure.  For test  purposes, we also
can consider  the flow to  be polytropic.   We have verified  that for
those models that  are provided as examples in this  paper, the effect
of the  adopted equations of state  on mass flux does  not exceed $\pm
4\%$.

We  also  should  mention  that  in  our scheme  we  do  not  use  any
approximation for  an optically  thin MT  stream that  originates from
above the  photosphere.  The primary reason  for this is that  we have
not  yet developed  a  model  that would  allow  one  to combine  both
optically thick and  optically thin ML schemes when  a donor overfills
its Roche lobe. \cite{KolbRitter1990} have suggested that during RLOF,
the  total MT  rate  can be  found  by a  summation of  the  MT of  an
optically thick stream, calculated as  a function of the current RLOF,
plus the maximum  MT rate obtainable for an  optically thin photospheric
flow in the case of RL underflow, which is at the instant when a star 
exactly fills its Roche  lobe.  We  are not  confident whether  this method
would   estimate  correctly   the   contribution   of  an   isothermal
photospheric  outflow in the case of  a non-negligible  RLOF, considering
that both the geometrical cross-section and  the potential of  the nozzle
for  the  isothermal flow  (located  now  around  the nozzle  for  the
adiabatic  flow)  are  changed   substantially  from  those  in the $L_1$
neighbourhood. Hence, we do not use this approach.  

The degree to which optically thin mass transfer affects the behaviour of
stars before their photospheres overfill the Roche lobe depends on the rate
of their intrinsic evolutionary expansion when their photospheres approach the Roche lobe.
Relatively massive giants (5~$\msun$ and up),
cross the interval of radii for which the optically thin mass transfer is dominant
in a short time thanks to their fast expansion. Hence, the fraction of the envelope they lose via optically thin transfer is small.
On the other hand, less massive giants epxand slowly and can lose a lot of mass through atmospheric ML before the optically
thick mechanism kicks in. This decreases the mass ratio at the onset of RLOF and improves stability.
We observed an extreme case, where with the \cite{Ritter1988} prescription, a $0.89~\msun$  giant could never actually overfill a
$100\rsun$ Roche lobe, and steadily lost the whole convective envelope via optically thin mechanism reaching an optically thin MT rate of $10^{-3.6}~\msun \rm{yr}^{-1}$!
We therefore warn the reader that the mode of mass transfer from low-mass giants might be sensitive to the very uncertain model of optically thin MT.

We summarise  the simplifications eliminated in the  existing optically
thick schemes and in our case in Table \ref{tab:opt_thick}.

\begin{table}
\caption{Simplifications eliminated in the optically thick mass transfer schemes}
\begin{tabular}{lcccc}
\hline
Reference & GS  & PC & PD & PS  \\
\hline

\cite{Paczynski1972}   &   &  &  &  \\
\cite{Savonije1978}    &   &  &  &   \\
\cite{KolbRitter1990}  &   &  & $\bullet$ &   \\
\cite{Ge10}   &  &  $\circ$ & $\bullet$ &    \\
This work   & $\bullet$ & $\bullet$ & $\bullet$ & $\bullet$ \\

\hline
\end{tabular}
\label{tab:opt_thick}
\medskip

GS -- geometrical simplification for the nozzle, PC -- pressure correspondence, PD -- polytropic stratification of the donor, PS -- along the streamlines $P/\rho^{\gamma}$ is constant.
Empty circle -- simplified potential is adopted the effect of which is equivalent to the PC assumption.
\end{table}

\section{Results of ML simulations in red giants} \label{paper_sim_results}

\subsection{Response to constant ML} \label{paper_sim_results_const}

\begin{figure}
\includegraphics[width=88mm]{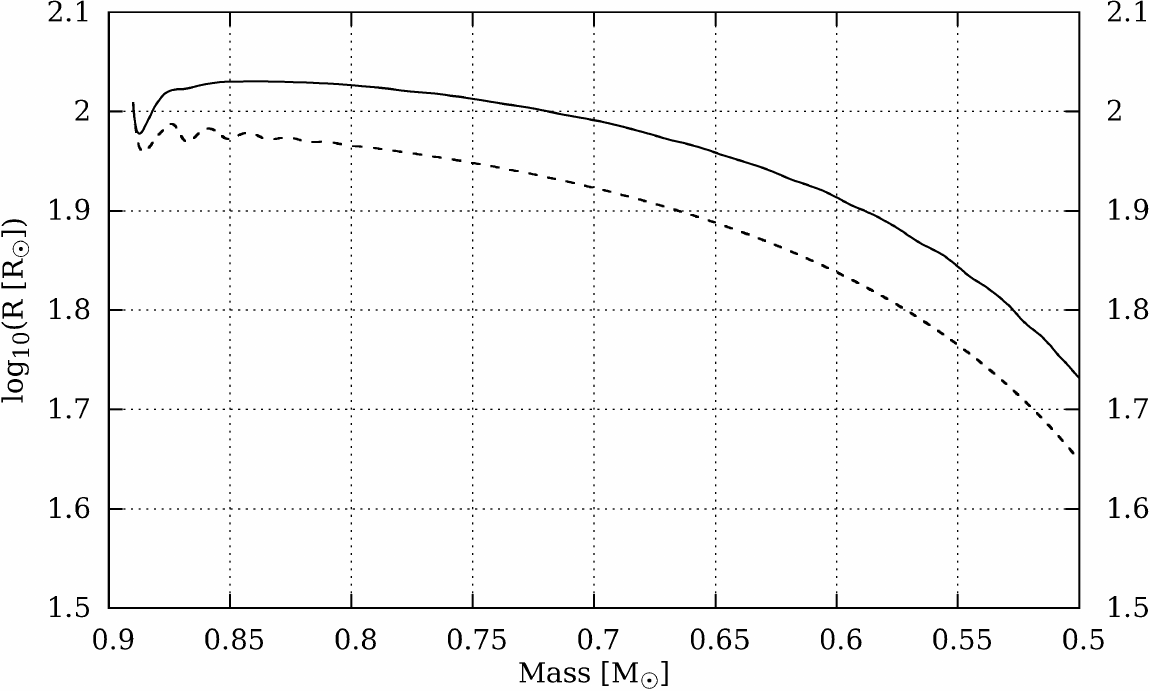}
\caption{
Models passy-a and passy-b. A $102~\rsun$, $0.89~\msun$ red giant 
similar to the one considered by \protect\cite{Passy2012} is subjected to 
constant ML of $0.1~\msun$~yr$^{-1}$. 
The solid line represents the result we reproduced following the method 
described by \protect\cite{Passy2012}, which is very similar to the one shown 
by a red line in their Figure 4. The dashed line represents the result 
we obtained by compensating for the limitations discussed in
 Section~\ref{paper_massloss_eps_g}, 
and after removing some pulsation artefacts by disabling composition 
smoothing at the bottom of the convective envelope.}
\label{fig:passy}
\end{figure}

First, we can compare the initial  reaction of a red giant to constant
ML to  our prediction made  in Section~\ref{paper_massloss_hresponse}.
For example, we take a $5M_\odot$ and $50R_\odot$ giant and subject it
to the  constant ML rate  $\dot M = 10^{-3.5}  M_\odot$~yr$^{-1}$. The
initial       derivative       of      radius       predicted       by
formula~(\ref{eq:consump_der_red})    is    about   $    -1.7    \cdot
10^{5}~\rm{cm\ s^{-1}}$. With modified  boundary conditions and with a
timestep   of   $10^{-4}$~yr,   the    code   produces   $-1.5   \cdot
10^{5}~\rm{cm\ s^{-1}}$.  If instead  the default hydrostatic boundary
condition    is    used,    the     code    produces    $-1.2    \cdot
10^{5}~\rm{cm\ s^{-1}}$.

Second,  to compare our results  to  the ones  published earlier,  we
calculate the behaviour of the $0.89~\msun$ and $102 R_\odot$ red giant
examined by \cite{Passy2012}.  As in \cite{Passy2012}, we subject this
giant to  the constant ML  rate $\dot  M = 0.1  M_\odot$~yr$^{-1}$. We
evolve  one giant using the unmodified {\tt MESA} code with viscosity
and timestep setup described in \cite{Passy2012}; we call this the ``passy-a'' model.
For the other giant, ``passy-b'' model, we take into account the circumstances
outlined      in      Section~\ref{paper_massloss_eps_g}.
In   particular,  to   obtain   correct
$\epsilon_{\rm  g}$, we evolve this model using a constant time  step of 
$10^{-5}$~yr.  We  also remove some pulsation  artefacts by disabling
composition smoothing at the bottom of the convective envelope.

The initial behaviour of the radius  in both models has been explained
in Section~\ref{paper_massloss_hresponse}.  Note, however, that in our
model,   the  stellar   radius  is   always  smaller   than  that   in
\cite{Passy2012} (see our  Figure~\ref{fig:passy} and their Figure~4).
We find that the difference in the value of the radii is mainly due to
how accurately the value of $\epsilon_{\rm g}$ is found.

In addition to the initial  radius contraction, one can notice ensuing
radial  oscillations,   visible  both   in  the  models   passy-a  and
passy-b.  We define  those  non-numerical  radial pulsations,  excited
either by the start of constant ML  or, in the case of evolution in a
binary,  by the  rapid growth  of the  ML rate,  as mass  loss induced
pulsations (MLIPs). These pulsations might be caused by the sonic
rarefaction wave, reflected from the bottom of the envelope, which is theoretically
predicted to occur in a fluid as it abruptly expands into vacuum,
hence we think that they are not of numerical nature.
In the model passy-a, these pulsations are largely
smoothed out because the timestep in  this model grows after the start
of simulations and exceeds the dynamical timescale.

A crucial  parameter that defines the  level of importance of  MLIP is
the  p-mode  damping  rate $\eta_{\rm{p}}$,  which  characterises  the
timescale on which a giant roughly attains hydrostatic equilibrium and
dynamical  oscillations are  damped. With  contemporary high-precision
photometric instruments  such as COROT  and Kepler, it is  possible to
obtain high-precision measurements of the profile widths (i.e, damping
rates  divided  by  $\pi$)   of  p-modes  \citep[see  \eg][]{Baudin11,
  Belkacem12}.  Damping rates  of intrinsic  pulsations of  red giants
found  from  those  observations are  $\eta_{\rm{p}}\sim10^{-1}\pi$  to
$10^{1}\pi$.

The damping rates, given that  the timestep used
is  much smaller  than  the  pulsation period,  depend  hugely on  the
artificial viscosity coefficients: increasing the artificial viscosity
increases  the damping  rates. The  damping  rates of  MLIPs shown  in
Figure~\ref{fig:passy} are  comparable, within an order  of magnitude,
with those obtained directly from  observations. Note that the default
artificial viscosity used  for the models passy-a and  passy-b is very
moderate  ($l_{1} =  0.1$, $l_{2}  =  0$) and  does not  substantially
affect  the damping  rates. Due  to the lack  of wide-range  observational
calibrations for  damping rates across  giants of different  radii and
masses, we do not use an artificial viscosity in our models, except in
models passy-a  and passy-b, where it  is taken into account  only for
the purpose of comparison with the original paper of \cite{Passy2012}.

We have  performed several  simulations where  we subjected  giants of
several  initial masses  and radii  to constant  ML, to  determine the
realistic stellar response defined as:

\begin{equation}
 \zeta \equiv  \left ( \frac{\partial \log R}{\partial \log M} \right)  \ .
\end{equation}
\noindent  Note that generally $\zeta$ is an implicit function of not 
only the current MT rate, but also of the previous MT, and here we 
look at $\zeta$ for constant MT rates only. We found that:

\begin{figure}
\includegraphics[width=84mm]{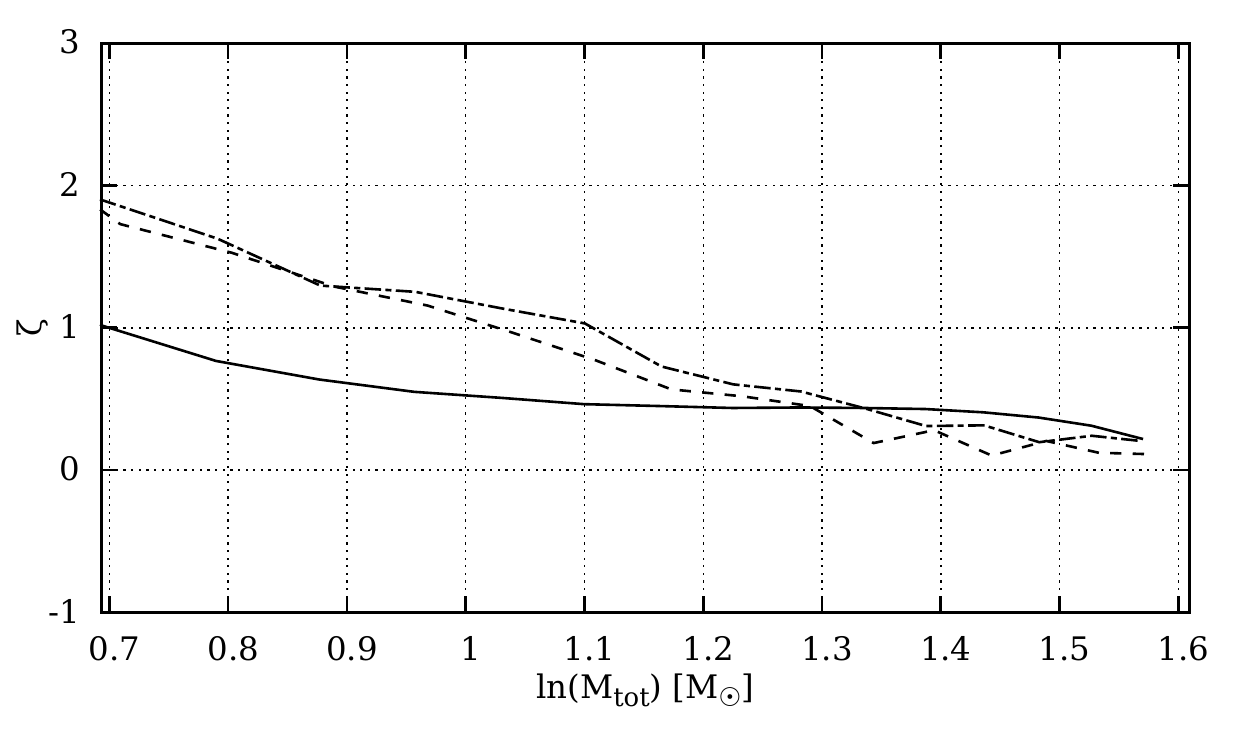}
\caption{$\zeta$ for a giant of initial mass 5.0~$\msun$ and radius 50~$\rsun$ 
subjected to various rates of mass loss ($10^{-3}$, $10^{-2}$ and $10^{-1}$~$\msun$~yr$^{-1}$
(solid, dashed and dot-dashed lines respectively).}
\label{fig:zeta_var}
\end{figure}

\begin{figure}
\includegraphics[width=84mm]{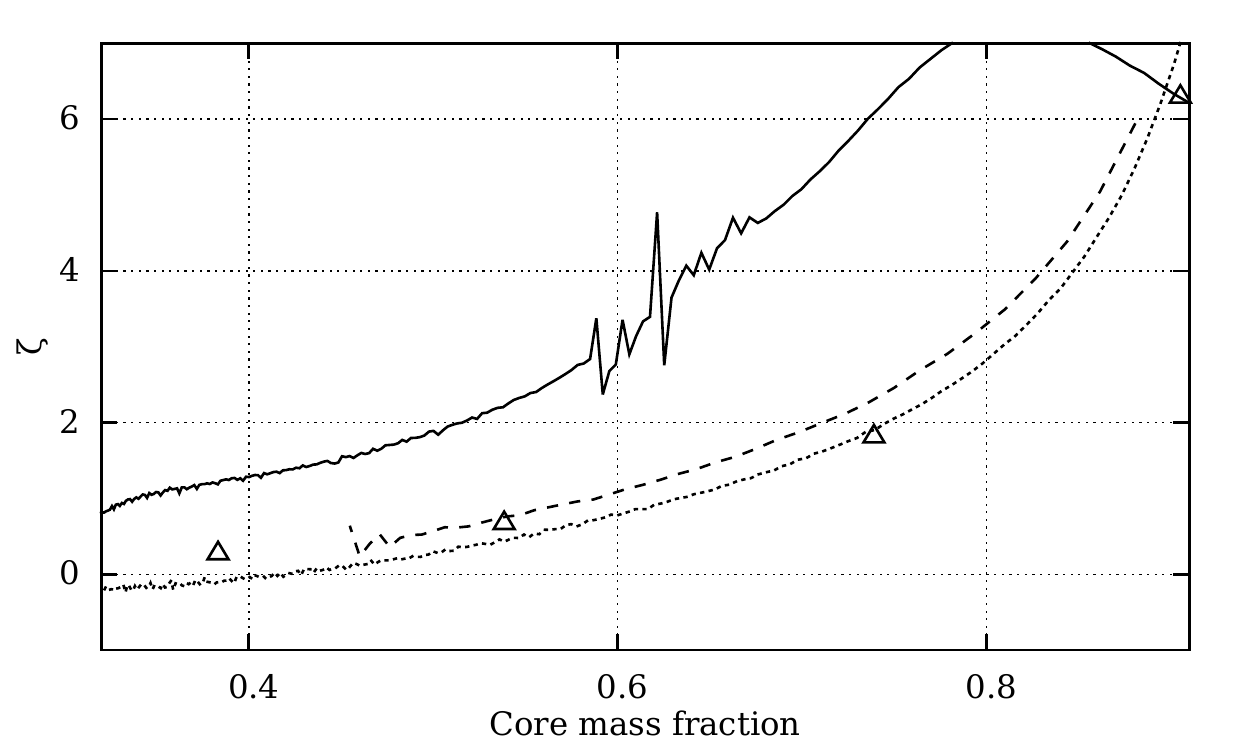}
\caption{$\zeta$      for      giants      at     MT      rate
  $10^{-1}~\msun\rm{yr^{-1}}$. 5.0  $\msun$ 50~$\rsun$  (solid line),
  1.0  $\msun$ 100~$\rsun$  (dashed line),  5.0 $\msun$  200~$\rsun$
  (dotted  line). $\zeta_{\rm  comp}$ for  a composite  polytrope ($n  =
  3/2$),    taken    from    \protect\cite{Hjellming87},   is    shown    with
  triangles.  Initial  portions  of   the  responses  with  irrelevant
  dynamical oscillations are removed.}
\label{fig:zeta_poly}
\end{figure}

\begin{enumerate}[(i)]

\item  At  high ML  rates,  $\zeta$  saturates.  It becomes  mainly  a
  function  of the  core mass  fraction and  not of  the ML  rate (see
  Figure~\ref{fig:zeta_var} where we show representative examples);

\item In  ``well-expanded'' giants,  the saturated  $\zeta$ approaches
  the  behaviour of  a  composite polytrope  $\zeta_{\rm  comp}$, as  in
  \cite{Hjellming87}, with $n = 3/2$ (see Figure~\ref{fig:zeta_poly}).
  In more compact giants, $\zeta$  is higher than what is expected
  from the composite polytrope.

\item  Non-saturated values  of  $\zeta$ are  usually  lower than  the
  saturated $\zeta$.

\item  $\zeta$ at the very onset of RLOF  is often  hard to  determine due  to
  MLIPs.

\end{enumerate}

While the radial  response seems to be most natural  to look at during
the  MT, we  find that  other properties of the donor's surface also depend
dramatically on how well $\epsilon_{\rm g}$ is obtained. In Section
2.5  we  showed  that  substantial  luminosity  can  be  generated  if
recombination  energy is  calculated properly.   We can compare the
surface luminosities of $5 M_\odot$  and $50 R_\odot$ giant, subjected
to a constant ML $\dot M =  0.1 M_\odot$~yr$^{-1}$ and evolved with a
default time-step  adjustment and with  a time-step chosen  to resolve
$\epsilon_{\rm g}$.  An estimate provided in Section~2.5 shows that
the energy coming from recombination in a giant subjected to ML is approximately

\begin{equation}
W_{\rm rec}\approx 10^{5} L_\odot \frac {\dot M}{M_\odot {\rm} yr^{-1}} \ .
\end{equation}

Where does this energy mainly go -- is it radiated away as excess luminosity
or spent on the star itself, \eg, to increase its thermal or gravitational energy?
\noindent  The unperturbed  giant  has luminosity  only about  $L_{\rm
  donor}\sim  10^{3} L_\odot$.  The  calculations show  that a
mass-losing   giant   in   which   $\epsilon_{\rm   g}$   in   the
superadiabatic layer was calculated properly, has a luminosity about 6
times higher  than the  mass-losing giant  with a  default relatively
large time-step! The first giant has also become about 50\% hotter.
This increase in luminosity indicates that a large portion of energy
from recombination is radiated away, rather than being spent on the star itself.
In the case of a $5  M_\odot$  and $200  R_\odot$  giant,  a  simple prediction  for  a
luminosity increase and an exact calculation would agree fairly well -- in
a more expanded giant, less recombination energy was spent on the star
itself,  and most  went  directly to  surface luminosity.   Similarly, the 
luminosities in the passy-a and passy-b  models are different by  about 6
times, also  as a  simple estimate  would predict. Interestingly, in the 
case of $5 M_\odot$ giants, the radii obtained by the two methods
were not  much different, except that  in giants evolved with  a small
time-step we can observe MLIPs.

Of course, very fast MT rates are not commonly observed. However,
the path of  a star through the  fast MT is different,  and can define
both the stability and/or how the donor appears when the fast MT phase
is  completed.   We  also  can  evaluate that  this  effect  might  be
important (i.e. provides a difference in the luminosity by more than a
few per cent) for as long as

\begin{equation}
{\dot M} \ga 10^{-7} \frac{M_\odot}{\rm yr} \times \frac{L_{\rm donor}}{L_\odot} \,
\label{eq:mdotrec}
\end{equation}
\noindent and hence might be important for low-luminosity giant donors and, in general, 
for all donors transferring mass at the thermal timescale of their envelope.

Indeed, consider 
the thermal timescale $\tau_{\rm KH}$ of the envelope taken as usual 
\begin{equation}
\tau_{\rm KH}\approx 2\times 10^7\ {\rm yr} \left ( \frac{M}{M_\odot}\right )^2 \frac{R_\odot}{R} \frac{L_\odot}{L} \ . 
\end{equation}

\noindent For most giants $R/R_\odot\gg 2 M/M_\odot$, and  their envelope thermal timescale MT, $M/\tau_{\rm KH}$, satisfies condition \ref{eq:mdotrec}.


\subsection{Stability of the Roche lobe mass transfer} \label{paper_sim_results_scml}

The conventional way to determine stability of a binary system to
the MT is to compare the initial responses of the donor radius and
the Roche lobe radius to the mass loss.  As was mentioned in Section~1,
such analysis, if it does not involve actual MT calculations in a binary, is
usually done by comparing their mass-radius exponents at the onset of RLOF.
In order to do this, the adiabatic radial response of the donor $\zeta_{\rm ad}$
is found from one of the existing approximations (composite polytropes, condensed
polytropes, or even simply adopting that for convective donors
$\zeta_{\rm ad}=0$). This adiabatic radial response is used instead of
the realistic stellar response.

However, we'd like to stress, that even if the realistic
stellar response is obtained, there is no reason to  
assume that if $\zeta < \zeta_{\rm{L}}$ at the start of the MT, and a
donor's relative Roche lobe overflow {\it initially} increases,
the instability will necessarily occur. Instead of considering just the moment of RLOF,
we can trace binary evolution through the MT in
detail and determine if the instability eventually takes place, or
not. The determination of MT instability in a binary system, while the
donor is overflowing its RL, requires an alternative (to a
simple comparison of the initial $\zeta$s)
criterion to delineate when the MT
proceeds stably, and when it results in a common envelope.

To understand the characteristic behaviour of the systems during MT, we performed a
large set of simulations with donors of several initial masses
($1,2,5,10,15, 30, 50 M_\odot$). We considered several values of radii for each
donor, taken within the range where the donors have 
non-negligible, $\ga 0.3 M_{\rm{donor}}$ outer convective envelopes,
both at the red giant and asymptotic giant branches.
We conducted simulations for several initial mass ratios
(between 0.9 and 3.5). We assume that a companion is compact, where the compactness only means
that we neglect any possible accretor's RLOF.

Further, for these detailed MT sequences we adopted fully
conservative MT; any non-conservation in the MT would lead to a
relative increase of MT stability. Most of our binary systems were
evolved using our modifications: obtaining proper $\epsilon_{\rm g}$, and our method to find MT rates and boundary
conditions.
 
\begin{figure}
\includegraphics[width=84mm]{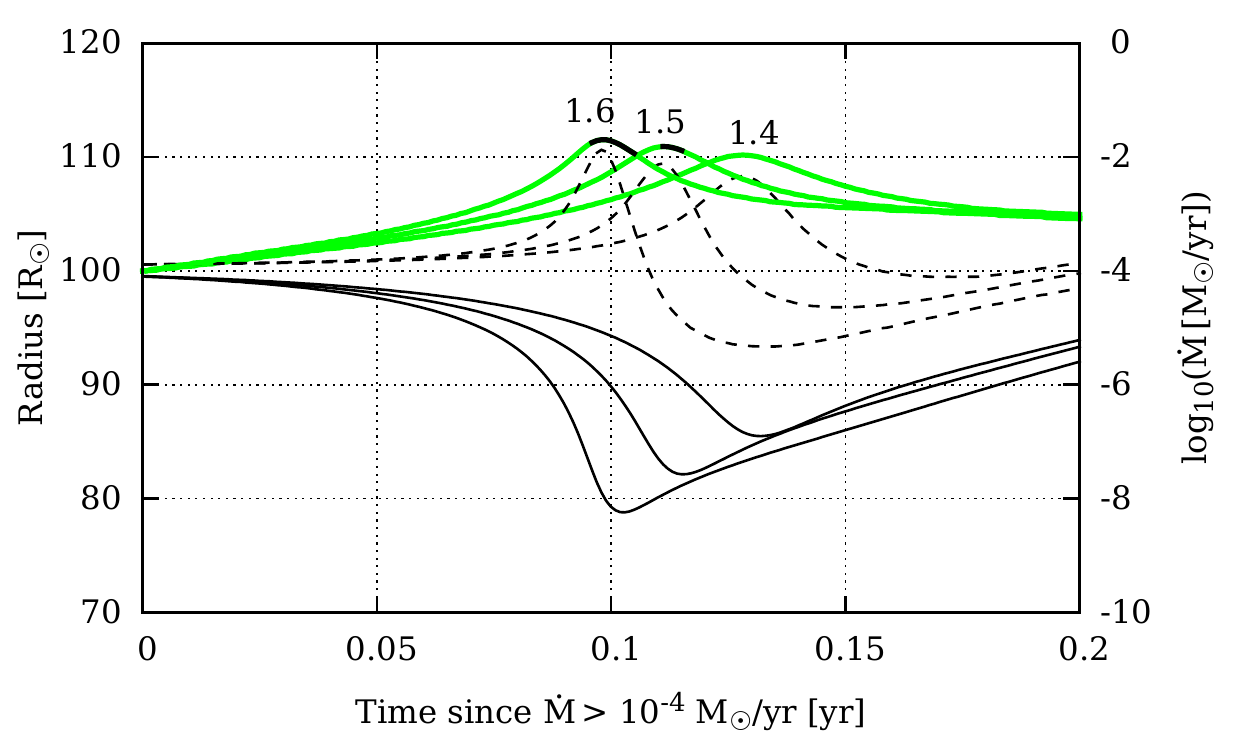}
\caption{The MT rate (green solid lines), and the evolution of the donor's 
radius (dashed lines) and the donor's Roche lobe radius (solid black lines), 
shown for the mass ratio with which the binary system does not start 
$L_2/L_3$-overflow during the MT ($q=1.4$), and when $L_2/_L3$ 
takes place ($q=1.5, 1.6)$. The period of $L_2/L_3$ overflow is indicated with solid 
black colour on the MT curves. The donor is a $1 M_\odot$ 
and $100 R_\odot$ giant at the onset of the MT. 
The rate of a dynamical MT for this system is $\sim 10\, M_\odot$\ yr$^{-1}$. }
\label{fig:stab_l2l3}
\end{figure}

Even for systems with larger initial $q$ than that dictated by the
conventional adiabatic stability criterion, the MT rate after RLOF smoothly increases, approaches a peak
value and then decreases; the peak MT rate is usually much less than
the dynamical MT rate.  With the increase of $q$, the peak MT rate
increases. We define as $q_{L23}$ the smallest initial mass ratio for which the
binary experiences $L_2/L_3$-overflow during the MT (see Figure \ref{fig:stab_l2l3}).  We refer to this overflow as 
$L_2/L_3$-overflow, as during the MT, the mass ratio can reverse, and
$L_3$-overflow becomes $L_2$-overflow. This $q_{L23}$ is a function of
both the initial mass ratio and the donor's radius.  For many initial
donor radii, in binary systems with $q=q_{L23}$, the mass of the
donor encompassed between the donor's Roche lobe and its $L_2/L_3$ lobe
is very small and the ML rate when $L_2/L_3$-overflow is approached is
far from dynamical. Numerically, we are capable of evolving such systems
with $q>q_{L23}$ through the  $L_2/L_3$-overflow, and many of them will
not even approach the dynamical MT rate (see Figure \ref{fig:stab_l2l3}).
We, however, do not trust the MT rates obtained in this regime because our stream model only considers
the $L_1$-nozzle.

We adopt therefore that for as long as
$L_{\rm{2}}/L_{\rm{3}}$-overflow of the donor does not happen, the
system is stable.  In part, this is confirmed by the three-dimensional
hydrodynamical simulations, in which a common envelope is always
associated with a severe, albeit very short in duration,
$L_{\rm{2}}/L_{\rm{3}}$ overflow of the donor
\citep{2014ApJ...786...39N}.  Note however, that we consider this to
be a {\it minimum} stability criterion, as the MT in a binary system
may remain stable even after the donor's $L_{\rm{2}}/L_{\rm{3}}$-overflow
takes place, as, e.g., SS 433 shows \citep[e.g.,][]{2010A&A...521A..81B,2010MNRAS.408....2P}.

We also compared the surface-averaged effective gravitational acceleration over the $L_2/L_3$ lobe
with the spherically symmetric acceleration that would be expected at its volume radius and found
that the difference is about 13\%. 

In addition to $L_{\rm{2}}/L_{\rm{3}}$-overflow condition, we trace
whether the binary orbit is changing rapidly.  For the latter
 condition, we adopt that if $|\dot a/a| T < 1/50$, the orbit is not changing rapidly, where $T$ is the binary period.
Rapid evolution of orbital parameters can invalidate the entire Roche formalism. For example, if
angular velocity changes substantially over one period, the Euler term of the fictitious force, which is ignored in
the Roche formalism along with the Coriolis term, becomes comparable to the centrifugal term.
Note, however, that in Nature,
binary systems with a larger $|\dot a/a| T$ might not experience
dynamical-timescale MT.

\begin{figure}
\includegraphics[width=84mm]{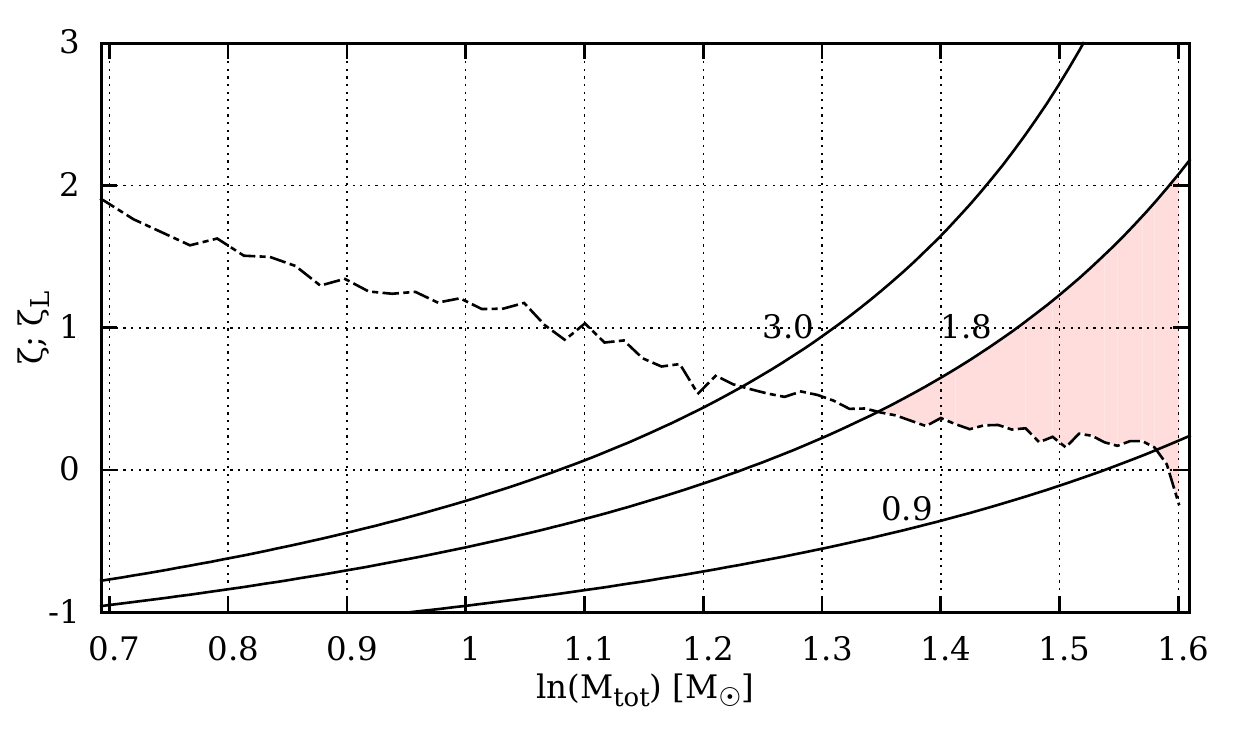}
\caption{$\zeta$ for a giant of initial mass 5.0~$\msun$ and radius 50~$\rsun$ 
subjected to a constant mass loss of $10^{-1}$~$\msun$~yr$^{-1}$ (dot-dashed line).
Solid lines show $\zeta_{\rm{L}}$ for the conservative MT at various initial 
mass ratios.
Intersections $\zeta = \zeta_{\rm{L}}$ define the critical mass-points 
$m_{\rm cp}$, at which mass transfer 
starts to decrease. The shaded area is equal to  
the right part of the Equation \ref{eq:area} for $q=1.8$ and $m = m_{\rm cp}$.}
\label{fig:zeta}
\end{figure}

\begin{figure}
\includegraphics[width=84mm]{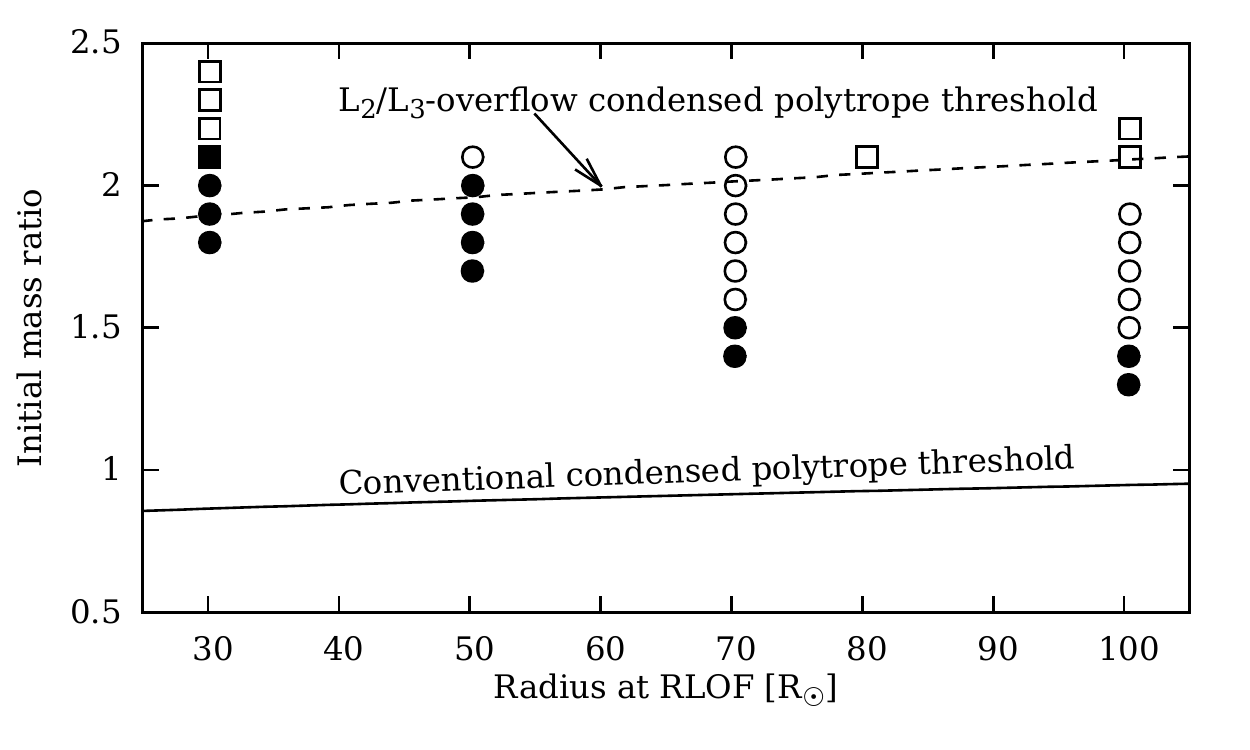} 
\includegraphics[width=84mm]{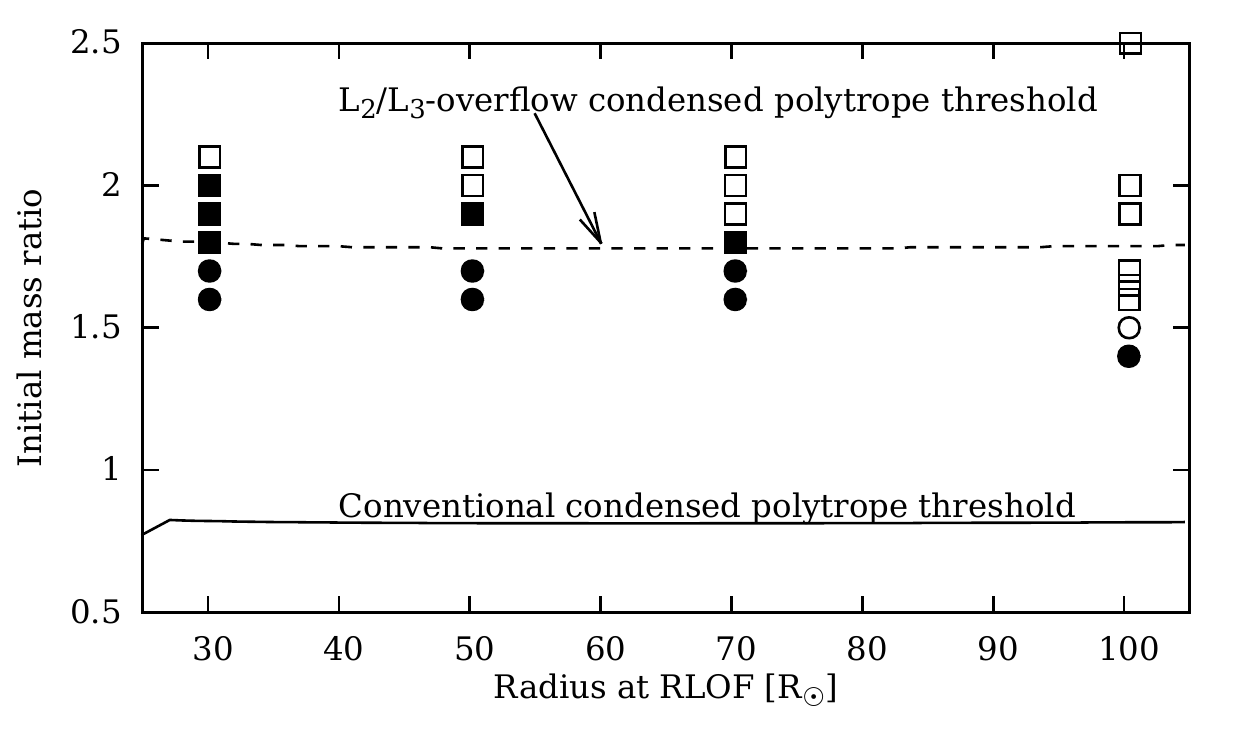}
\includegraphics[width=84mm]{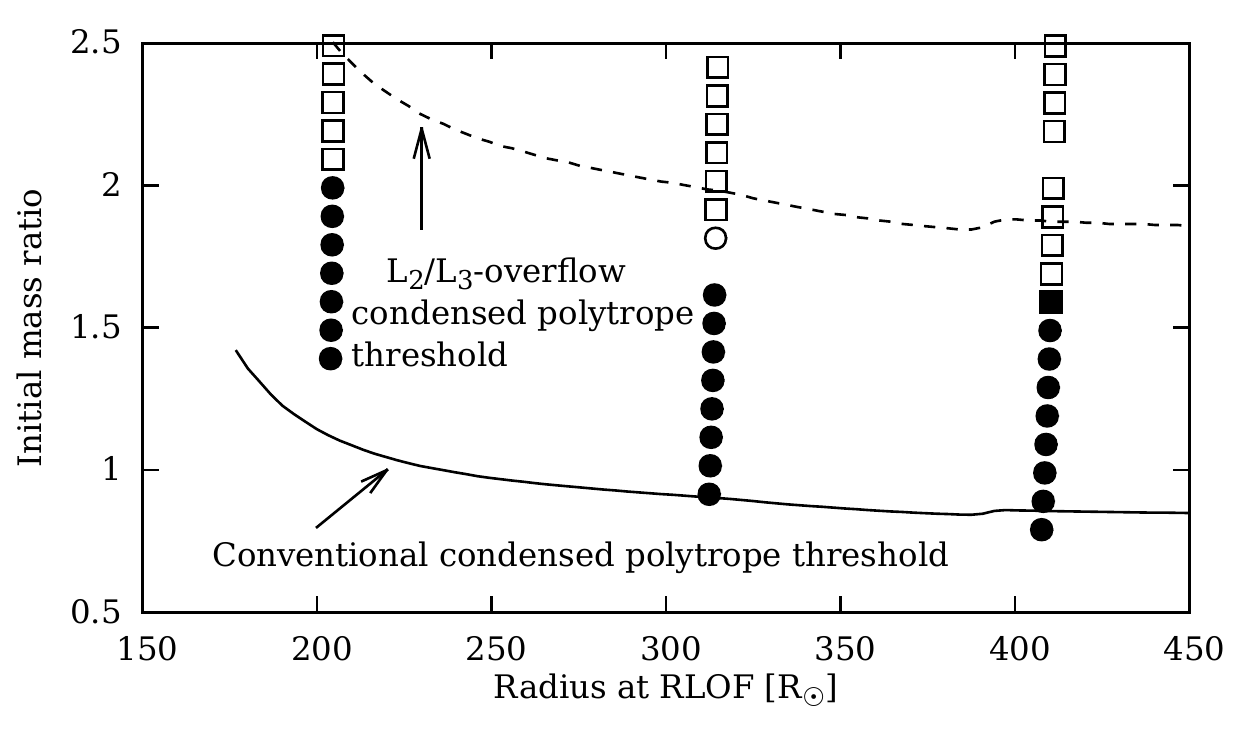}
\caption{ The outcomes of the MT sequences for donors of different initial masses.
The top panel is for a $1 M_\odot$ donor, the middle panel is for a $2 M_\odot$ 
donor and the bottom panel is for a $10 M_\odot$ donor. A square symbol indicates 
that $|\dot a|T/a > 1/50$, otherwise, the symbol is a circle. If a symbol 
is filled, then no $L_2/L_3$ overflow is experienced, if a symbol is empty, then the system experiences
$L_2/L_3$-overflow.  Our $L_2/L_3$-overflow condensed polytrope 
simplification is shown with a dashed line, and $q_{\zeta}$ 
 -- the conventional condensed polytrope threshold -- is shown with a solid line.}
\label{fig:stab}
\end{figure}

The key reason behind the unexpected stability of the MT in systems
with $q_{\zeta} < q < q_{L23}$ is that once the MT starts,
$\zeta$ {\it rises} as the donor's mass decreases (see
Figures~\ref{fig:zeta_var}, \ref{fig:zeta_poly} and \ref{fig:zeta}).
At the same time, $\zeta_{\rm L}$ decreases (Figure \ref{fig:zeta}). A
decrease in $(\zeta_{\rm L}-\zeta)$ means that, as the MT proceeds,
the stability of the system increases.  We define as the
\emph{critical mass-point}, $m_{\rm cp}$, the mass of the donor at
which $\zeta = \zeta_{\rm L}$.  When the donor decreases its mass
to the critical mass-point, 
the ratio of the donor radius to its Roche lobe 
 starts to decrease.  Therefore, assuming that the relative RL overflow
at the $L_{\rm{2}}/L_{\rm{3}}$ equipotential weakly depends on the mass ratio (see below),
if there was no $L_{\rm{2}}/L_{\rm{3}}$ before the donor mass decreased to $m_{\rm cp}$, the
binary system is stable with respect to MT.
We have searched for the point when $L_{\rm{2}}/L_{\rm{3}}$-overflow starts
using our extended set of detailed simulations (see Figure
\ref{fig:stab}).  It can be seen that the critical mass ratios
$q_{L23}$ for which the MT becomes ``unstable'' are about twice as large 
as would be predicted by conventional comparison of
mass-radius exponents at the onset of the MT.

Let us  analyse the  results and show  how the  approximate critical
mass   ratios  can   be  predicted   without  doing   detailed  binary
calculations.

To  predict  whether   a  system  experiences
$L_{\rm{2}}/L_{\rm{3}}$-overflow during the MT, we  need  to
identify whether at any moment after the start of MT the donor's radius exceeds $R_{\rm  L23}$ -- the
volume    radius     at    which    the     donor    starts
$L_{\rm{2}}/L_{\rm{3}}$-overflow.   In order to do this, we  can find  the
ratio $R_{\rm  L23}/R_{\rm L}$ for a range of the donor masses. Note that this critical ratio also depends on the mass ratio.
However,  we find that this dependence is quite weak, $\ln (R_{\rm  L23})/R_{\rm  L}\approx  0.27\pm0.01$ for  $0.7\le
q\le 4$. Nevertheless, we took this dependence into account in our calculations by examining the ratio $R_{\rm  L23}/R_{\rm L}$ not only down to 
donor mass $m_{\rm cp}$, but also further, almost all the way down to the donor core mass. 

The radius of the donor during the MT,  when it has shed to mass $m$, is

\begin{equation}
\ln R (m) = \ln R_{0} + \int_{M_0}^m \zeta(\dot M) dm \ .
\label{eq:radzeta}
\end{equation}

It follows from Section 5.1 that in order to  predict the radius 
of the donor at any moment during fast (saturated) MT,
instead  of the real $\zeta$ one can use  $\zeta_{\rm  comp}$, which is obtained from
the composite polytrope approximation. ``Compact''  giants, should  be expected  to
expand  less than  this  estimate  would predict,  and  hence be  more
stable because their $\zeta>\zeta_{\rm comp}$.

Using   the   definitions   of   mass-radius   exponents   (see   also
Figure~\ref{fig:zeta}), we can  find the approximate  radius of the
donor at any moment after the start of MT as

\begin{equation}
\ln \left( \frac{R(m)}  {R_{\rm L}} \right) = \int_{M_{0}}^{m} \left( \zeta_{\rm comp} - \zeta_{\rm L} \right) d \ln M \ .
\label{eq:area}
\end{equation}

\noindent Now we can find whether the system experiences $L_2/L_3$-overflow at any
moment during the MT and thus produce a semi-analytical estimate of $q_{L23}$ -- we will refer to this estimate as 
``$L_2/L_3$-overflow condensed polytrope simplification''. It is plotted in Figure~\ref{fig:stab} for various giants.

We can see that the condensed polytrope simplification works best for
those giants that are neither too compact nor too expanded.  In both
compact and average-sized giants that approach
$L_{\rm{2}}/L_{\rm{3}}$-overflow at the critical point, most mass
before the critical point is lost at ML rates which are generally much
higher than $\sim 10^{-2}~\msun~{\rm yr^{-1}}$, so it's fine to use the saturated value of $\zeta$.
At the same time, compact giants have higher saturated
values of $\zeta$, which makes them more stable than average-sized
giants whose saturated $\zeta$ approximately follows the condensed
polytrope.

In large giants with very rarefied envelopes,
$L_{\rm{2}}/L_{\rm{3}}$-overflow occurs at much lower mass loss rates,
and the saturated values of $\zeta$ become inapplicable.  Instead, a
lower, non-adiabatic, non-saturated values should be used.  This makes
these giants less stable. However, such $L_2/L_3$-overflow will be non-dynamical
and does not have to result in a common envelope.

\begin{figure}
\includegraphics[width=84mm]{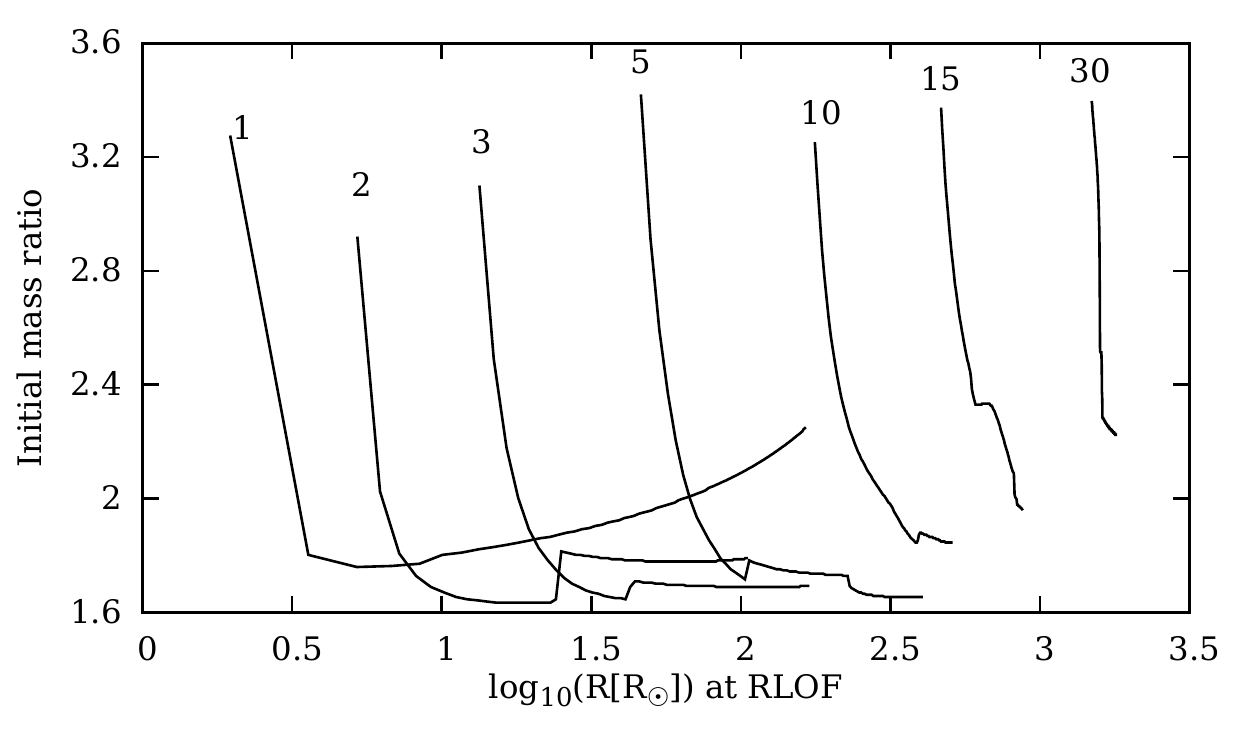}
\caption{ Critical mass ratios obtained from the $L_2/L_3$-overflow simplified 
composite polytropes for the donors of different initial masses and radii.}
\label{fig:qcrit}
\end{figure}

In Figure~\ref{fig:qcrit} we show how the critical mass ratio obtained 
from the $L_2/L_3$-overflow condensed polytrope
simplification changes with the initial donor mass and radius. 
We have also checked and found 
that non-conservative MT leads, as expected, to the increase
in $q_{L23}$, increasing it by about 0.3.

The detailed simulations (Figure~\ref{fig:stab}) show that $q_{L23}$ 
is larger when a donor is more compact and the mass fraction 
of its convective envelope is smaller. 
In all detailed simulations we did with donors 
where the outer convective envelope is non-negligible, $q_{L23}$ is below 3.5, 
where 3.5 is known as the critical mass ratio leading to a delayed 
dynamical instability with radiative donors  \citep[see, e.g,
][]{Ge10}. 
It shows therefore that during the development of the outer
convective envelope, the convective donors are in the transitional
regime.   When the convective envelope is well developed, the critical
mass ratios are in the range $1.5-2.2$ for most donors. 

We conclude that the criterion based on $L_{\rm{2}}/L_{\rm{3}}$ overflow is definitely
predicting that more binary systems will evolve through their MT in a stable way.

Whether the $L_{\rm{2}}/L_{\rm{3}}$ overflow should necessarily lead to a common envelope
is not entirely clear. In principle, MT through the $L_{\rm{2}}/L_{\rm{3}}$ nozzle can be
treated in the same simplified way as that through the $L_{1}$ nozzle, and the material can be considered
to leave the system carrying away the specific angular momentum of the donor. It is a subject of our future work.

\section{Conclusions} 

\label{paper_conclusion}

In this paper, we considered a number of theoretical and practical
challenges on the path to understanding the stability of the MT in
binaries consisting of a convective giant donor and a compact
accretor.

We find that in order to obtain the correct response to the mass loss, the 
$\epsilon_{\rm grav}$ in the superadiabatic layer must be calculated
properly; depending on which stellar code is used this may require a
number of numerical tricks, and at the very least  control of the 
time-step to match the predicted energy generation rate. We provide
simple estimates for how to find the mass of this superadiabatic layer
and the rate of the energy release, in order to quantify how well an
arbitrary stellar code performs under mass loss.

We have shown that the mass-radius exponents in the giants that are compatible
with the composite-polytrope description converge to the prediction in
\cite{Hjellming87} for fast mass loss. On the other hand, the giants
that cannot be described by a composite polytrope have
$\zeta>\zeta_{\rm comp}$, and the binary systems with such donors are
more stable than the composite polytrope would predict.
In our detailed models, we could not find the strong superadiabatic
expansion predicted by the adiabatic models in \cite{Ge10}.
 
In addition to radial response, we examined the changes in surface
luminosity and  effective temperature of mass-losing stars. We
find that at fast ML rates, the luminosity can differ by a factor of several
times between the models where $\epsilon_{\rm grav}$ in the
superadiabatic layer is calculated correctly and the ones where it is
calculated in the Lagrangian way with large timesteps.
We note that in some observed binary systems, the donor's properties
such as effective temperature and mass are determined with a few \%
precision \citep[e.g.,][]{2014ApJ...793...79L}.  This difference in
effective temperature obtained by the two models, and comparable to the observed
precision, is expected to occur in systems with a low-luminosity giant
donor, and also in thermal-timescale MT systems. 

We have enhanced the classical scheme to find the MT rate via an 
optically thick stream approximation. In particular, our use of the 
detailed system geometry allows us to find the MT rate in the case of a
substantial RLOF. This leads us to a new criterion of MT instability,
based on whether the donor starts $L_{\rm{2}}/L_{\rm{3}}$ overflow. We
have also found that a binary system does not become immediately
unstable as the donor's envelope becomes convective, but rather 
the mass ratio at which the instability occurs gradually
decreases from the regime predicted
for radiative donors. Note that in principle a binary can survive the
MT even after $L_{\rm{2}}/L_{\rm{3}}$ overflow, without starting a
common envelope.  However, we find that even the new
$L_{\rm{2}}/L_{\rm{3}}$-overflow criterion warrants that binary
systems will proceed with stable conservative MT if their mass ratio is
up to twice that given by the conventional criterion, for stars with deep 
convective zones (and the mass ratio can be even larger for shallow outer convective zones).

\section*{Acknowledgements}
K.P. acknowledges support from Golden Bell Jar fellowship.
N.I. acknowledges support by NSERC Discovery Grants and the Canada Research Chairs Program;
this research was supported in part by the National Science Foundation under Grant No. NSF PHY05-51164.
Authors thank B.~Paxton for help with MESA/star and MESA/eos modules, T.~Fragos for help with MESA/binary module, C.~Heinke for help with the manuscript and the anonymous referee
for useful comments.
Authors are also grateful to R.~Webbink, Ph.~Podsiadlowski and S.~Justham for useful discussions.
This research has been enabled by the use of computing resources provided by WestGrid and Compute/Calcul Canada.

\bibliographystyle{mn2e}
\bibliography{mt}

\end{document}